\documentclass[aps,prb,showpacs,twocolumn]{revtex4}
\usepackage{amsmath,graphics}
\usepackage[next]{inputenc}
\usepackage[dvips]{epsfig}
\def\be{\begin{equation}}
\def\ee{\end{equation}}

\def\bi{\begin{itemize}}
\def\ei{\end{itemize}}
\def\bn{\begin{enumerate}}
\def\en{\end{enumerate}}
\def\bea{\begin{eqnarray}}
\def\eea{\end{eqnarray}}
\def\no{\nonumber}
\def\ba{\begin{array}}
\def\ea{\end{array}}
\def\bd{\begin{displaymath}}
\def\ed{\end{displaymath}}

\begin{document}
\title{Finite temperature topological order in 2D topological color codes}

\author{Mehdi Kargarian}
\email[]{kargarian@physics.sharif.edu}

\affiliation{Physics Department, Sharif University of Technology,
Tehran 11155-9161, Iran}

\begin{abstract}
In this work the topological order at finite temperature in
two-dimensional color code is studied. The topological entropy is
used to measure the behavior of the topological order. Topological
order in color code arises from the colored string-net structures.
By imposing the hard constrained limit the exact solution of the
entanglement entropy becomes possible. For finite size systems, by
rising the temperature one type of string-net structures is
thermalized and the associative topological entropy vanishes. In the
thermodynamic limit the underlying topological order is fragile even
at very low temperatures. Taking first thermodynamic limit and then
zero-temperature limit and vice versa don't commute, and their
difference is related only to the topology of regions. The
contribution of the colors and symmetry of the model in the
topological entropy is also discussed. It is shown how the gauge
symmetry of the color code underlies the topological entropy.

\end{abstract}
\date{\today}

\pacs{03.67.Lx, 03.67.Mn, 03.65.Ud}

\maketitle
\section{Introduction \label{introduction}}
Exotic sates of matter are those that defy the usual description in
terms of well-known Landau-Ginzburg-Wilson paradigm
\cite{goldenfeld}, where an appropriate local order parameter
characterizes different behaviors of two phases on either side of
the critical point. These are new phases of matter carrying a kind
of quantum order called \emph{topological order}\cite{wen1} and the
transition between various phases does not depend on the symmetry
breaking mechanism. From the experimental point of view, the
fractional quantum Hall liquids \cite{wen2, read, froh} exhibits
topological order. Different phases of electron liquids carry the
same symmetry and the phase transition between them is possible.
Therefore, the Landau theory of classical phase transitions fails in
order to describe these phases. The most remarkable properties of
such new states are the dependency of the ground state degeneracy on
genus or handles of the space, gapped excitations and non-trivial
braiding of excitations. This latter property introduces new class
of emerging particles with the statistical properties that are
neither fermions nor bosons. In fact, they are \emph{anyons}. They
interact topologically independent of their distances much like that
of Aharonov-Bohm interaction. If by winding of an anyon around
another one the wavefunction picks up an overall phase, the anyons
are called \emph{abelian} anyons. But, if evolution of the
wavefunction is captured by a unitary matrix, the anyons are called
\emph{non-abelian} anyons, i.e. they obey non-abelian braiding
statistics.

The questions such as what the objects underlying the topological
phases are and how different phases can be classified are still
under debate. However, some physical mechanisms such as
string/membrane-net condensation\cite{wen3,wen4},
branyons\cite{miguel1} that are analogous of the particle
condensation in the symmetry-breaking phases and a description in
terms of quantum groups\cite{bais} have been introduced. The
string-nets are extended nonlocal objects and the ground state is a
coherent superposition of all possible string's configurations
appearing at all length scale. This physical picture clarifies the
topological order based on the microscopic degrees of freedom.
Emerging particles such as fermions or anyons are collective
excitations of strings\cite{wen5}.

The fact that in topological ordered phases the ground state
subspace has a robust degeneracy and the excitations above the
ground state are gapped give them the ability of being rigorous
quantum memory\cite{dennis} in the sense of error correction.
However, these are not sufficient conditions for being
self-correcting code. Perhaps thermal noises spoil the
self-stability of the code\cite{Alicki,Nussinov}. The construction
of fault-tolerant topological quantum computation\cite{kitaev1}
exploits the emerging properties of topological ordered phases. The
quantum information can be stored on the topologically protected
subspace being free from the
decoherence. 
The robust manipulation of quantum information is done by braiding
of anyons\cite{Freedman1,Freedman2,Nayak,Pachos}, where the unitary
gate operations are carried out by braiding of anyons.

The ground state of topological ordered phases is highly entangled
state and an indicator for topological order which is not based on
symmetry has been emerged, notably the topological entanglement
entropy\cite{kitaev2,levin}. This topological quantity appears as
subleading correction to the entanglement entropy of a convex
region. It is a general feature of all gapped phases that the
entanglement entropy scales with the boundary of the region, the so
called \emph{area law}\cite{plenio}. However, the appearance of
constant subleading term is a new feature related to topological order. 
The topological phase can be described as a phase of matter for
which the low-energy effective theory is a topological field theory
(TQFT). Topological entanglement entropy is related to one of the
basic parameter of the TQFT, the total quantum dimension of emerging
quasiparticles in the theory.

The best known model for studying topological order, emerging
abelian or non-abelian anyons and examining its capabilities for
topological quantum computations is the famous Kitaev's
model\cite{kitaev3}. In the abelian phase, the model becomes the
well-known toric code with a stabilizer structure\cite{gottesman}.
The stabilizer structure of the code is given by a set of
\emph{star} and \emph{plaquette} operators. Stabilizer operators fix
a subspace in the Hilbert space of the model in which different
states are distinguished by means of topological numbers.
\emph{Topological color code} is another relevant example of
stabilizer codes, with enhanced computational
capabilities\cite{hector1, hector2}. The stabilizer operators are
only colored plaquette operators. However, both codes are
topological because the stabilizer operators are local and
non-detectable errors have non-trivial supports on the manifold.
Non-trivial string operators stand for encoded logical operators.
Topological order in toric code and color code are related to
different gauge symmetry. Indeed, the topological order in the toric
code is related to the $\mathbf{Z_{2}}$ gauge symmetry, whereas the
topological order in the color code is related to the
$\mathbf{Z_{2}}\times\mathbf{Z_{2}}$ gauge symmetry. This latter
symmetry arises from the contribution of the colors in the
construction of the code and is responsible of the string-net
structure of the code. However, both models have almost the same
error threshold when one considers the error syndrome
measurements\cite{miguel2}.

Entanglement properties of both models at zero-temperature have been
extensively studied. The entanglement entropy of a region with its
complement depends on the degrees of freedom living on the boundary
of the region supplemented with a topological term. But, the
topological term for the color code\cite{kargarian} is just twice
than the toric code\cite{hamma} when both lattices are put on a
compact surface. The difference arises because the excitations in
the toric code are ascribed to the four superselection sectors,
whereas the contribution of the colors in the color code model makes
the excitations richer than those of the toric code. In fact in the
color code the number of superselection sectors is sixteen.

How much topological order is robust against the thermal
fluctuations? This is a very important question from the practical
point of view since every practical implementation of quantum
information processing can not ignore the effects of thermal
fluctuations. Thermal fluctuations may create excitations (errors)
accumulating in the system and destroying the quantum information
encoded in the states. Original idea of topological order which is a
zero temperature concept can be extended to thermal mixed
states\cite{claudio1}, where classical limit arises by washing out
the off diagonal elements of the density matrix. With such
realization of classical topological order, it would be tempted to
investigate the topological order in terms of temperature. It can be
used as a measure of resilience of a code against the temperature.
We would like to point out at finite temperature the entanglement
entropy is no longer a measure of correlations between subsystems.
Instead, the mutual information does as it scales with the boundary
of subsystems. The topological order however manifests itself as
subleading terms of entanglement entropy as well as mutual
information. As long as we are interested in the topological order,
a linear combination of entropies or mutual
information\cite{pachos1} of subsystems gives a measure of
topological order, since in the combination the dependency on bulk
and boundary degrees of freedom are washed out. In the toric code
each of its underlying loop structure or gauge structure contributes
exactly half to the topological entanglement entropy. This is
expected since both loop structures are similar in a fashion that
they are defined on the original lattice and its dual. For a finite
size system topological order survives even at non-zero
temperatures\cite{claudio2,pachos1}. Further increasing of
temperature destroys the loop structures of the model implying the
fragility of the toric code \cite{Alicki,Kay,Nussinov2}.

In this work we address the above problem, i.e the fate of
topological order at finite temperature, in the color codes. The
loop structures of the color code is different from that of toric
code since they are related to different gauge symmetries. We
discuss how gauge symmetry affects the finite temperature properties
of the code. We attach to each set of plaquettes with the same color
an energy scale. In the lattice gauge theory these energy scales are
translated into the chemical potential for creating the respective
charges. Following the derivation of C. Castelnovo, \emph{et
al}\cite{claudio2}, we first impose the hard constrained limit on
the string-net structure in $\sigma^{z}$ bases and allow for the
thermalization of the string-net structure in $\sigma^{x}$ bases.
Then, in order to identify the contribution of colored strings in
the topological entanglement entropy, we fix other loop structures
and examine the residual topological order. Also in the high
temperature limit the description in terms of classical topological
order is recovered.

The organization of the paper is as follows. In the next section the
color codes is briefly reviewed. Then in Sec(\ref{matrix}) the
thermal density matrix that is needed for subsequent arguments is
derived. In Sec(\ref{entanglement entropy}) the entanglement entropy
is derived from the density matrix. Then, limiting behavior of the
entanglement is discussed in Sec(\ref{limiting}). Topological
entanglement entropy and its behavior in terms of temperature is the
subject of Sec(\ref{topological}). The case of open boundary
conditions is discussed in Sec(\ref{triangular}).
Sec(\ref{conclusion}) is devoted to conclusions.

\begin{figure}
\begin{center}
\includegraphics[width=8cm]{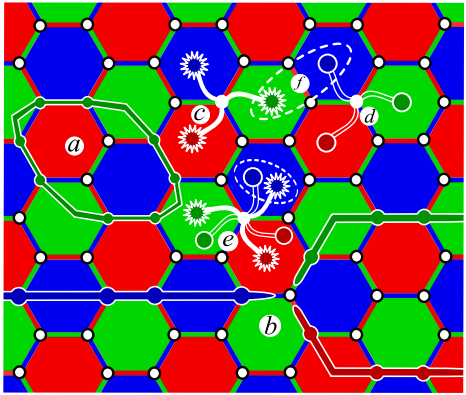}
\caption{(color online) A piece of 2-colex with colored hexagons as
its plaquettes. Lightest to darkest hexagons: green, red and blue,
respectively (a) The closed green string corresponds to the product
of blue and red plaquettes. (b) A generalization of the closed
colored strings, the string-net. (c) Performing a $\sigma^{z}$
rotation creates excitations with topological charges $(e,\chi_{c})$
(d) $\sigma^{x}$ rotation  creates other types of excitations with
charges $(c,\chi_{e})$ (e) $\sigma^{y}$ rotation creates both types
of quasiparticles at corresponding plaquettes. The composite
particle in the dashed ellipse is a boson with charge $(b,\chi_{b})$
(f) The composite particle in the dashed ellipse is a fermion with
charge $(b,\chi_{g})$.} \label{fig1}
\end{center}
\end{figure}

\section{2-colex and color code: fixing notations \label{color code}}
Let us start by a brief introductory on the color code model.
Consider a two-dimensional trivalent lattice composed of plaquettes,
vertices and links. Such structure is shown in Fig.\ref{fig1}. To
keep track of vertices, plaquettes and links, we use color as a
bookkeeping tool. We will use three colors: red, green and blue.
Then we color the plaquettes of the lattice in which two neighboring
plaquettes don't share in the same color. In this way we can also
color links so that a $c$-link (the letter $c$ stand for color
throughout the paper unless stated another meaning) connects two
$c$-plaquettes. We call this two-dimensional lattice a
\emph{2-colex}. The lattice can also be embedded in higher spatial
dimensions, the so-called \emph{D-colex}\cite{miguel1,miguel3}.
Hence, the thermal analysis presented here can be extended to higher
dimensional cases as in the toric code\cite{claudio3}. Without loss
of generality we suppose all plaquettes are hexagons. Suppose the
2-colex is embedded on a compact surface such as torus, i.e imposing
periodic boundary conditions. To connect the lattice to a physical
system, we place a qubit at each vertex . With each plaquette we
attach two operators aimed at reaching the stabilizer code. These
operators are simply the product of a set Pauli operators acting on
the qubits around a $c$-plaquette: $X_{c}=\prod_{i\in
\partial P}\sigma_{i}^{x}$ and $Z_{c}=\prod_{i\in \partial
P}\sigma_{i}^{z}$, where $\partial P$ stands for six qubits around a
$c$-plaquette. With this definition for plaquette operators we see
that all plaquettes of either color are treated on equal footing.
Although the colors play no role in the plaquette operators, they
determine the structure of the coding space. It is simple to see
that all plaquette operators commute with each other since they
share either nothing or at even number of qubits. The plaquette
operators are generators of a stabilizer group; thereby there are
vectors in a subspace of the whole Hilbert space that are common
eigenvectors of all plaquette operators with eigenvalue $+1$. This
subspace, the coding space, is spanned by some vectors that are
distinguished by eigenvalues of a set of non-local operators in
which their supports wind the handles of the
space\cite{hector1,kargarian}. Although three distinct colors
introduce the code, there is an interplay between color and homology
of the space. In fact for each homology class only two colors are
independent related to the symmetry of the code. As long as we
consider closed spaces such as torus, the closed non-contractible
loops are enough to form bases for the coding space. The coding
space is called \emph{color code}. From now on, unless it is stated
else, we suppose the 2-colex spanned by $3N$ plaquettes, $N$ of each
color. Notice that on a compact surface like a torus all plaquette
operators are not independent. In fact they are subject to the
constraint $\prod_{g\in \mathcal{G}}\Omega_{g}=\prod_{b\in
\mathcal{B}}\Omega_{b}=\prod_{r\in \mathcal{R}}\Omega_{r}$, where
$\Omega=X,Z$, and $\mathcal{G}$, $\mathcal{B}$, $\mathcal{R}$ are
sets of green, blue and red plaquettes, respectively.

The protected subspace is ground state of a many-body Hamiltonian
that is minus sum of all plaquette operators equipped with some
coupling constants as follows:\bea \label{eq1}
H&=&-\lambda^{g}_{x}\sum_{g\in\mathcal{G}}X_{g}-\lambda^{b}_{x}\sum_{b\in\mathcal{B}}X_{b}-\lambda^{r}_{x}\sum_{r\in\mathcal{R}}X_{r} \nonumber \\
&&-\lambda^{g}_{z}\sum_{g\in\mathcal{G}}Z_{g}-\lambda^{b}_{z}\sum_{b\in\mathcal{B}}Z_{b}-\lambda^{r}_{z}\sum_{r\in\mathcal{R}}Z_{r},\eea
where $\lambda^{c}_{x,z}$'s are coupling constants and each sum runs
over all corresponding $c$-plaquettes. The ground states spanning
the coding space are all vectors in which
$X_{c}|\psi\rangle=Z_{c}|\psi\rangle=|\psi\rangle$ for all plaquette
operators. Any violation of this condition amounts to an excited
state or alternatively as an \emph{error}. Therefore, the ground
state is said to be vortex free in the sense of being closed
string-net condensate. In fact, the ground state is an equal
weighted superposition of all string-net configurations. Such
configurations can be visualized if we consider the product of
plaquette operators. For example consider the product of two
neighboring red and blue plaquettes. In the product, the Pauli
operators of qubits shared between two plaquettes square identity
and we are left with a green string being boundary of two
plaquettes. This closed green string has been shown in
Fig.\ref{fig1}(a), which connects green plaquettes. Such closed
strings of either size and color commute with Hamiltonian. Another
interesting extended object underlying the ground state structure is
string-net that is a collection of colored strings. A typical
string-net has been depicted in Fig.\ref{fig1}(b). The associated
string-net operator can be the product of either $\sigma^{x}$ or
$\sigma^{z}$ acting on the qubits it contains (filled colored
circles on the string-net). This string operator commutes with all
plaquettes since they share either nothing or even number of qubits.
The appearance of such string-net is crucial for the full
implementation of the Clifford group\cite{hector1}.

Excitations appear as end points of open colored strings. An open
string anticommutes with the plaquettes lying at its ends as they
share odd number of qubits. To have a simple picture of excitations,
let us focus on some simple cases. Consider a rotation $\sigma^{z}$
applied to a certain qubit as in Fig.\ref{fig1}(c). As $\sigma^{z}$
anticommutes with $X_{r}$, $X_{b}$ and $X_{g}$ plaquette operators
adjacent to the qubit, it will increase the energy of plaquettes by
$2\lambda^{r}_{x}$, $2\lambda^{b}_{x}$ and $2\lambda^{g}_{x}$,
respectively. In this case the excitation on a $c$-plaquette is
revealed by an $c$-star. Similarly, if we perform a $\sigma^{x}$
rotation on a qubit as in Fig.\ref{fig1}(d), the anticommutation of
$\sigma^{x}$ with $Z_{c}$ of neighboring plaquettes leads the energy
of red, blue and green plaquettes increase by $2\lambda^{r}_{z}$,
$2\lambda^{b}_{z}$ and $2\lambda^{g}_{z}$, respectively. In this
case the excitation on a $c$-plaquette is revealed by a $c$-circle.
If a rotation $\sigma^{y}$ is performed, as in Fig.(\ref{fig1}e),
each plaquette contains both above excitations, i.e. the excitations
on the $c$-plaquette increase the energy by
$2\lambda^{c}_{x}+2\lambda^{c}_{z}$. All quasiparticles are by
themselves boson. However, they may have mutual semionic statistics.
For instance, if a red particle of Fig.(\ref{fig1}d) go around a
green particle of Fig.(\ref{fig1}c) , the wavefunction picks up a
minus sign implying they are anyons. There are also composite
particles of two excitations as shown by white dashed ellipses. If
two excitations differ in both color and type, they form a fermion
as in Fig.\ref{fig1}(f). Otherwise, they form a boson as in
Fig.\ref{fig1}(e).

All above emerging particles that are the low energy excitations of
the model can be classified in terms of underlying gauge
group\cite{kargarian2,bais2}. Before that, let us make a convention
for colors which will be useful for subsequent discussions. We refer
to colors by a bar operation $\bar{c}$ that transform colors
cyclically as $\bar{r}=g$, $\bar{g}=b$ and $\bar{b}=r$. The elements
of the gauge group $\mathbf{Z_{2}}\times \mathbf{Z_{2}}$ are
$\{e,r,b,g\}$. Each excitation carries a topological charge. The
corresponding topological charge can be labeled by pair $(q,\chi)$,
where $q\in \mathbf{Z_{2}}\times \mathbf{Z_{2}}$ and $\chi$ an irrp
of this group. Therefore, there are nine bosons labeled by
$(c,\chi_{e})$, $(e,\chi_{c})$ and $(c,\chi_{c})$ and six fermions
$(c,\chi_{\bar{c}})$ and $(c,\chi_{\bar{\bar{c}}})$. Taking into
account the vacuum with trivial charge $(e,\chi_{e})$, color code
has sixteen topological charges or superselection sectors. Now, we
can label the excitations in Fig.\ref{fig1} regarding their
topological charges. The excitations in Fig.\ref{fig1}(c) and
Fig.\ref{fig1}(d) have $(e,\chi_{c})$ and $(c,\chi_{e})$ charges,
respectively, while the boson in Fig.\ref{fig1}(e) and fermion in
Fig.\ref{fig1}(f) have $(b,\chi_{b})$ and $(b,\chi_{g})$ charges,
respectively.
\section{thermal density matrix of color code \label{matrix}}
Zero-temperature entanglement of the color code\cite{kargarian} is
given by writing the ground state of the Hamiltonian in
Eq.(\ref{eq1}). Let $G$ denotes the group constructed by a set of
generators of the spin-flip plaquette operators, i.e $X_{c}$. By
starting from a polarized state, the ground state will be an equal
weighted superposition of all elements of the stabilizer group as
follows.\bea \label{eq2} |\psi\rangle=|G|^{-1/2}\sum_{h\in G}
h|0\rangle^{\bigotimes |V|}, \eea where $h$ is an element of the
stabilizer group, $|G|$ is the cardinality of the group, $|V|$ is
total number of vertices on the lattice and
$\sigma^{z}|0\rangle=|0\rangle$. Other ground states can be
constructed through the action of the non-local spin-flip operators
with non-trivial supports winding the handles of the space. They are
distinguished by different topological numbers. However, all of them
have the same entanglement properties and we consider only one of
them as above.

Topological order in the ground state of the color code Hamiltonian
arises from the condensed string-net structures, which are loops
without open ends. The ground state $|\psi\rangle$ has been written
in the $\sigma^{z}$ bases. In each term of the superposition a
collection of spins related to the group element $h$ have been
flipped. Spins on the corresponding string-net have negative
$\sigma^{z}$ component. In this string-net structure $Z_{c}=1$ for
all plaquettes. There are also same string-net structures if we
worked in the $\sigma^{x}$ bases preserving the $X_{c}=1$. This is
because the model is symmetric upon the exchange of the $\sigma^{z}$
and $\sigma^{x}$ operators on the plaquettes. Unlike the toric
code\cite{claudio2}, both loop structures in the color code exist on
the direct lattice since for each plaquette we attach two operators.
The thermal fluctuations can break the string-nets and left them
with open ends. That how much thermal effects are able to destroy
the topological order associated to a string-net structure depends
on the coupling constants in the Hamiltonian as well as size of the
system. Definitely, we calculate the thermal density matrix and the
topological entanglement entropy is used as a measure of topological
order. Since the ground state is fixed point of all plaquette
operators, its underlying topological order arises from coherent
superposition of two string-net structures, each related to one of
the bases. We impose a constraint in which the string-net structures
related to the $\sigma^{z}$ bases are preserved. The so called hard
constrained limit\cite{claudio1} can be recasted into the limit
$\lambda^{c}_{z}\rightarrow \infty$. One may expect that in this
limit the thermal fluctuations can only create excitations with
topological charges $(e,\chi_{c})$. We can describe all excited
states as $|\psi[\gamma^{z}]\rangle=\gamma^{z}|\psi\rangle$, where
$\gamma^{z}$ is any arbitrary product of $\sigma^{z}$ Pauli
operators. The density matrix $\rho=\exp(-\beta H)$ in the bases
spanned by all states $|\psi[\gamma^{z}]\rangle$ is diagonal.
However, things get more simplified if we work in the original bases
$h|0\rangle^{\bigotimes|V|}$. Let us for simplicity drop the
superscript $\bigotimes|V|$. In these bases the density matrix reads
as follows \bea \label{eq3}\rho=\frac{\sum_{h,\tilde{h}\in
G}\langle0|he^{-\beta H}h\tilde{h}|0\rangle
h|0\rangle\langle0|h\tilde{h}}{\sum_{h\in G}\langle0|he^{-\beta
H}h|0\rangle}. \eea Without loss of generality let
$\lambda_{z}=\lambda^{c}_{z}$ for all plaquettes and
$G_{x}=\sum_{g}X_{g}$, $B_{x}=\sum_{b}X_{b}$, $R_{x}=\sum_{r}X_{r}$
and $Z=\sum_{c}Z_{c}$. Then the Hamiltonian proposed in
Eq.(\ref{eq1}) can be written as \bea \label{eq4}
H=-\lambda^{g}_{x}G_{x}-\lambda^{b}_{x}B_{x}-\lambda^{r}_{x}
R_{x}-\lambda_{z}Z.\eea Since $[Z,h]=0$ for all $h$, then
$Zh|0\rangle=3Nh|0\rangle$ and the numerator of Eq.(\ref{eq3}) reads
as\bea \label{eq5} \langle0|he^{-\beta
H}h\tilde{h}|0\rangle=e^{3\beta
\lambda_{z}N}\langle0|e^{\beta\lambda^{g}_{x}G_{x}+\beta\lambda^{b}_{x}B_{x}+\beta\lambda^{r}_{x}
R_{x}}\tilde{h}|0\rangle.\eea Since each plaquette operator squares
identity, i.e. $X^{2}_{c}=1$, each exponential term can be expressed
in terms of plaquette operators as follows. \bea \label{eq6}
e^{\beta\lambda^{g}_{x}G_{x}}=\prod_{g\in
\mathcal{G}}\left[\cosh(\beta\lambda^{g}_{x})+\sinh(\beta\lambda^{g}_{x})X_{g}\right].\eea
The two remaining exponential terms $e^{\beta\lambda^{b}_{x}B_{x}}$
and $e^{\beta\lambda^{r}_{x}R_{x}}$ have the same expressions and
obtained by replacing the color index $g$ by $b$ and $r$,
respectively. Every element of the stabilizer group $\tilde{h}$ has
a decomposition in terms of plaquette operators that it involves as
$\tilde{h}=\prod_{g\in \tilde{h}}X_{g}\prod_{b\in
\tilde{h}}X_{b}\prod_{r\in \tilde{h}}X_{r}$, where by $c\in
\tilde{h}$ we imply if the $c$-plaquette operator exists in the
decomposition of $\tilde{h}$. Let $n_{c}$ be the number of
$c$-plaquettes that exist in the decomposition of $\tilde{h}$.
Hence, the only nonzero terms in Eq.(\ref{eq5}) are as follows. \bea
\label{eq7}
&&\langle0|...|0\rangle=(\cosh\beta\lambda^{g}_{x})^{N}(\cosh\beta\lambda^{b}_{x})^{N}(\cosh\beta\lambda^{r}_{x})^{N}\times
\nonumber \\
&&\left[\sum_{c}e^{-(k_{\bar{c}}+k_{\bar{\bar{c}}})N-k_{c}n_{c}+k_{\bar{c}}n_{\bar{c}}+k_{\bar{\bar{c}}}n_{\bar{\bar{c}}}}+e^{-k_{g}n_{g}-k_{b}n_{b}-k_{r}n_{r}}\right],
\nonumber \\ \eea where $k_{c}=-\ln(\tanh\beta\lambda^{c}_{x})$ and
the sum runs over three colors $c\in \{r,b,g\}$. Notice that the
operator $\tilde{h}$ not only given by the product of green, blue
and red plaquettes belonging to $\tilde{h}$ but also by other
combinations of plaquettes due the periodic boundary conditions. In
fact, the operator $\tilde{h}$ has the following expressions in
terms of plaquettes operators.
\bea \label{eq8} \tilde{h} &=& \prod_{g\in
\tilde{h}}X_{g}\prod_{b\in \tilde{h}}X_{b}\prod_{r\in
\tilde{h}}X_{r}\no\\&=&\prod_{g\in \tilde{h}}X_{g}\prod_{b\in
\mathcal{B}\backslash \tilde{h}}X_{b}\prod_{r\in
\mathcal{R}\backslash \tilde{h}}X_{r}\no\\&=&\prod_{g\in
\mathcal{G}\backslash \tilde{h}}X_{g}\prod_{b\in
\tilde{h}}X_{b}\prod_{r\in \mathcal{R}\backslash
\tilde{h}}X_{r}\no\\&=&\prod_{g\in \mathcal{G}\backslash
\tilde{h}}X_{g}\prod_{b\in \mathcal{B}\backslash
\tilde{h}}X_{b}\prod_{r\in \tilde{h}}X_{r}. \eea 
Note that in the above expressions by $c\in \mathcal{C}\backslash
\tilde{h}$ we mean a $c$-plaquette in $\mathcal{C}$
$(\mathcal{C}=\mathcal{R},\mathcal{G},\mathcal{B})$ that is not in
$\tilde{h}$. Now we can see that the ambiguity in Eq.(\ref{eq7})
amounts to the above expressions for $\tilde{h}$. The partition
function can be evaluated in a similar way as follows:
 \bea \label{eq9} \mathcal{Z}&=&e^{3\beta
\lambda_{z}N}(\cosh\beta\lambda^{g}_{x})^{N}(\cosh\beta\lambda^{b}_{x})^{N}\times
\nonumber \\
&&(\cosh\beta\lambda^{r}_{x})^{N}\left(1+\sum_{c}e^{-(k_{\bar{c}}+k_{\bar{\bar{c}}})N}\right).\eea
Thus, the density matrix can be recasted into the following form:
\bea \label{eq10} \rho=\frac{1}{|G|}\sum_{h,\tilde{h}\in
G}\eta_{T}(\tilde{h})\times h|0\rangle\langle0|h\tilde{h},\eea where
\bea \nonumber &&\eta_{T}(\tilde{h})=
\no\\&&\frac{\sum_{c}e^{-(k_{\bar{c}}+k_{\bar{\bar{c}}})N-k_{c}n_{c}+k_{\bar{c}}n_{\bar{c}}+k_{\bar{\bar{c}}}n_{\bar{\bar{c}}}}+e^{-k_{g}n_{g}-k_{b}n_{b}-k_{r}n_{r}}}{\sum_{c}e^{-(k_{\bar{c}}+k_{\bar{\bar{c}}})N}+1}.
\nonumber\eea
The limiting behavior of the density matrix at zero and very high
temperatures is compatible with the respective known results. As
temperature tends to zero $(k_{c}\rightarrow 0)$, the pure density
matrix
$\rho=\frac{1}{|G|}\sum_{h,\tilde{h}}h|0\rangle\langle0|h\tilde{h}$
is recovered\cite{kargarian}. At high-temperature limit, the totally
mixed state $\rho=\frac{1}{|G|}\sum_{h}h|0\rangle\langle0|h$ yields
the classical limit of the model upon the hard constrained
limit\cite{claudio1}.
\section{entanglement entropy \label{entanglement entropy}}
\begin{figure}
\begin{center}
\includegraphics[width=8cm]{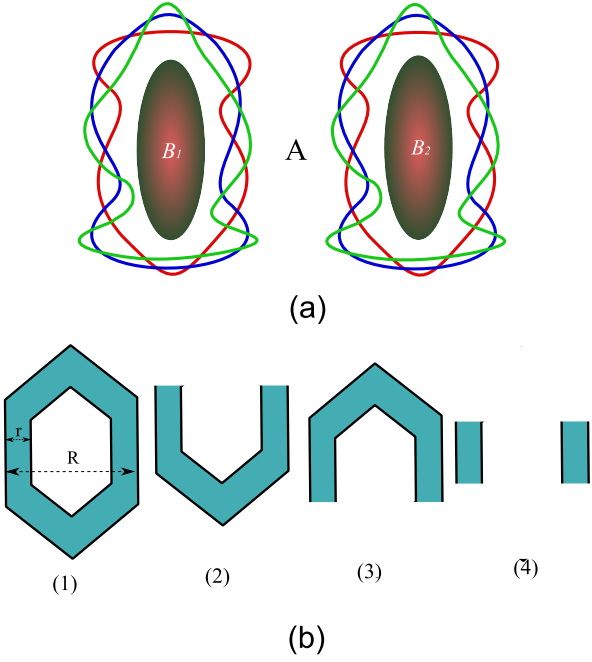}
\caption{(color online) (a) A schematic representation of simple
connected region $A$ and two disjoint regions $B_{1}$ and $B_{2}$.
For each disjoint region we can realize three colored collective
strings: blue(dark gray), red(gray) and green(light gray). All of
them are not independent as in text. (b) A set of bipartitions of
the lattice in different manner in which all bulk and boundary
contribution are dropped through the definition of the topological
entropy.} \label{fig2}
\end{center}
\end{figure}

In order to calculate the entanglement entropy we consider a generic
bipartition of the system into subsystems $A$ and $B$. Notice that
each bipartition may be composed of several disconnected regions.
Suppose $m_{A}$ and $m_{B}$ stand for the number of respective
bipartitions. Let $\Sigma_{A}$ and $\Sigma_{B}$ be the number of
plaquette operators acting solely on $A$ and $B$, respectively, and
let $\Sigma_{AB}$ stands for the number of plaquette operators
acting simultaneously on $A$ and $B$, i.e boundary operators. We
focus on the entanglement entropy between two partitions $A$ and $B$
of the system. To this end, first the reduced density operator of
the one subsystem is evaluated and then the entanglement entropy is
measured using von Neumann entropy. The reduced density matrix of a
region, say $A$, is obtained by tracing out $\rho$ with respect to
degrees of freedom of the subsystem $B$. Using the properties of the
group\cite{hamma}, the reduced density matrix reads \bea
\label{eq11} \rho_{A}=\frac{1}{|G|}\sum_{h\in G,\tilde{h}\in
G_{A}}\eta_{T}(\tilde{h})h_{A}|0_{A}\rangle\langle_{A}0|h_{A}\tilde{h}_{A},\eea
where $G_{A}$ and $G_{B}$ are subgroups of $G$ which act trivially
on subsystems \emph{B} and \emph{A}, respectively. The complete
description of the subgroup $G_{A}$ is given by a set of plaquette
operators acting solely on $A$ as well as collective operators. The
latter is a collection of operators acting solely on $A$, but they
are not the product of plaquettes in $G_{A}$. Suppose $A$ is a
simply connected region and the set $\{B_{1},B_{2},...,B_{m_{B}}\}$
presents the disconnected components of $B$. Let $AB_{i}$ be the
collection of plaquette operators acting simultaneously on $A$ and
$B_{i}$. The product of $c$- and $\bar{c}$-plaquettes of component
$B_{i}$ with the $c$- and $\bar{c}$-plaquettes of the boundary
$AB_{i}$ produces a $\bar{\bar{c}}$-string acting solely on $A$.
Denoting this string by $\gamma_{i}^{\bar{\bar{c}}}$, its expression
will be \bea \label{collective}
\gamma_{i}^{\bar{\bar{c}}}=\prod_{c,\bar{c}\in B_{i}\cup
AB_{i}}X_{c}X_{\bar{c}}. \eea A schematic representation of these
collective strings is shown in Fig.(\ref{fig2}a). Therefore, for
each disconnected region we can realize three collective operators
$\gamma_{i}^{r}$, $\gamma_{i}^{b}$ and $\gamma_{i}^{g}$. For each
disconnected region however only two of them are independent due to
the local constraint $\gamma_{i}^{r}\gamma_{i}^{b}\gamma_{i}^{g}=1$.
There is also another constraint on the total number of collective
operators. It is a global constraint: for a given color, say $c$,
the product $\prod_{i}\gamma_{i}^{c}$ can be produced by product of
all $\bar{c}$-and $\bar{\bar{c}}$-plaquettes acting solely on $A$,
namely $\prod_{i}\gamma_{i}^{c}=\prod_{\bar{c},\bar{\bar{c}}\in
A}X_{\bar{c}}X_{\bar{\bar{c}}}$. Taking into account all plaquettes,
collective operators and constraints on them, the cardinality of the
subgroups $G_{A}$ and $G_{B}$ will be
$d_{A}=2^{\Sigma_{A}+2m_{B}-2}$ and $d_{B}=2^{\Sigma_{B}+2m_{A}-2}$,
respectively.

The von Neumann entropy as a measure of entanglement between two
bipartitions is given by $S_{A}=-\mathrm{Tr}(\rho_{A}\ln\rho_{A})$.
We don't use this relation to measure the entanglement. Here, we
find it useful to instead compute it using the replica
$S_{A}=\lim_{n \rightarrow 1}
\partial_{n} \mathrm{Tr}[\rho^{n}_{A}]$. Hence, the trace of \emph{n-th} power of the reduced
density matrix will be\cite{claudio2} \bea
\label{eq12}\mathrm{Tr}[\rho^{n}_{A}]=\left(\frac{d_{B}}{|G|}\right)^{n-1}\prod^{n}_{l=1}\sum_{\tilde{h}_{l}\in
G_{A}}\eta_{T}(\tilde{h}_{l})\langle0|\tilde{h}_{1,A}\tilde{h}_{2,A}...\tilde{h}_{n,A}|0\rangle.
\nonumber \\ \eea In order to bring the above relation to a
manageable form, we resort to a map between the group elements
$\tilde{h}$ and the Ising variables\cite{claudio2}. Via this map the
group elements are labeled by Ising variables based on which
plaquettes and/or collective operators are involved in its
expression. let $\theta_{c}=-1(+1)$ be a Ising variables related to
appearing(not appearing) a $c$-plaquette in $\tilde{h}$. Similarly,
$\Theta^{c}_{i}=-1(+1)$ related to appearing(not appearing) a
collective $c$-string $\gamma^{c}_{i}$ in $\tilde{h}$. Notice that
this map is not a one-to-one map. In fact, it is a four-to-one map.
This latter point arises from the fact that by considering three
colors, if we reverse the sign of Ising variables related to two
colors, they represent the same element in the group. This point is
consistent with Eq.(\ref{eq8}) and will be clear in following. The
number of $c$-plaquettes, $n_{c}$, that the element $\tilde{h}$
involved can be expressed in terms of Ising variables as \bea
\label{eq13}n_{c}=\sum_{c\in
\mathcal{C}}\frac{1-\theta_{c}}{2}+\sum_{i}(\Sigma^{c}_{B_{i}}+\Sigma^{c}_{AB_{i}})\frac{1-\Theta^{\bar{c}}_{i}\Theta^{\bar{\bar{c}}}_{i}}{2}.
\eea Here, Once the color index in $n_{c}$ was fixed, the first sum
runs over all $c$-plaquettes in
$\mathcal{C}(\mathcal{G},\mathcal{B},\mathcal{R})$ and
$\theta_{c}$'s take values of $-1(+1)$ as defined above accordingly,
and the $\Sigma^{c}_{B_{i}}$ is the number of all $c$-plaquettes in
$B_{i}$. let $\Sigma^{c}_{B_{i}}+\Sigma^{c}_{AB_{i}}=\Sigma^{c}_{i}$
and since $\Sigma^{c}_{A}+\sum_{i}\Sigma^{c}_{i}=N$, Eq.(\ref{eq13})
can be written as \bea \label{eq14}
n_{c}=\frac{N}{2}-\frac{1}{2}\sum_{c\in
\mathcal{C}}\theta_{c}-\frac{1}{2}\sum_{i}\Sigma^{c}_{i}\Theta^{\bar{c}}_{i}\Theta^{\bar{\bar{c}}}_{i}.\eea
As we stated above, for a disconnected region $B_{i}$, the strings
$\gamma^{r}_{i}$, $\gamma^{b}_{i}$ and $\gamma^{g}_{i}$ are not
independent. With such dependency, the appearance of
$\Theta^{\bar{c}}_{i}\Theta^{\bar{\bar{c}}}_{i}$ in above relation
is meaningful. To see this, consider the case in which both
$\bar{c}-$ and $\bar{\bar{c}}-$strings are present in the group
element. This, in terms of Ising variables, yields
$\Theta^{\bar{c}}_{i}=\Theta^{\bar{\bar{c}}}_{i}=-1$. But, the local
constraint on strings implies that the product of two strings yield
the $c-$string. Since there is not any $c$-plaquette in the
decomposition of a $c$-string, the number of $c$-plaquettes arising
from the collective operators will be zero.

In terms of Ising variables, the Eq.(\ref{eq12}) will become \bea
\label{eq15}
\mathrm{Tr}[\rho^{n}_{A}]=\left(\frac{d_{B}}{|G|}\right)^{n-1}\frac{1}{4^{n}}\prod^{n}_{l=1}
\sum^{const.}_{\{\theta^{(l)}_{c},\Theta^{(l)c}_{i}\}}\eta_{T}\{\theta^{(l)}_{c},\Theta^{(l)c}_{i}\}\eea
The coefficient $\frac{1}{4}$ comes form the $4-1$ mapping between
Ising variables and plaquettes . The sum runs over all possible
Ising variables subject to a constraint. This constraint arises by
requiring the non-zero value for
$\langle0|\tilde{h}_{1,A}\tilde{h}_{2,A}...\tilde{h}_{n,A}|0\rangle$
in Eq.(\ref{eq12}), which implies that the product of $\tilde{h}$'s
must be trivial to give a non-zero value for the expectation value.
The constraint can be applied by the following expression:
\begin{widetext}
\bea \label{eq16}
\sum_{I_{r},I_{b},I_{g}}\Bigg\{\prod_{i}\left[\delta\left(\prod^{n}_{l=1}\Theta^{(l)b}_{i}\Theta^{(l)g}_{i}-I_{r}\right)
\delta\left(\prod^{n}_{l=1}\Theta^{(l)r}_{i}\Theta^{(l)g}_{i}-I_{b}\right)\delta\left(\prod^{n}_{l=1}\Theta^{(l)r}_{i}\Theta^{(l)b}_{i}-I_{g}\right)\right]\times
\nonumber \\
\prod_{r}\delta\left(\prod^{n}_{l=1}\theta^{(l)}_{r}-I_{r}\right)
\prod_{b}\delta\left(\prod^{n}_{l=1}\theta^{(l)}_{b}-I_{b}\right)\prod_{g}\delta\left(\prod^{n}_{l=1}\theta^{(l)}_{g}-I_{g}\right)\Bigg\}\delta\left(I_{r}I_{b}I_{g}-1\right),\eea
\end{widetext}
where $I_{c}=\pm1$ subject to $I_{r}I_{b}I_{g}=1$ which is imposed
by the delta function in the second line of above expression. Notice
that three delta functions in the first line are not independent and
we could already drop one of them. However, we keep them to make the
expression more symmetric. In fact, once two of them are fixed, the
third one satisfied. This is just a reinterpretation of the local
constraint that for a disconnected region only two strings are
independent.

Regarding the map between Ising variables and plaquettes, we can
fully bring the $\eta_{T}$ in Eq.(\ref{eq10}) into an expression in
terms of Ising variables.  Using Eq.(\ref{eq14}), the $\eta_{T}$
function reads as follows:
\begin{widetext}
\bea
\label{eq17}\eta_{T}=\frac{1}{4Z_{0}}\sum_{J_{r},J_{b},J_{g}}\left\{\prod_{r}e^{\frac{k_{r}}{2}J_{r}\theta_{r}}\prod_{b}e^{\frac{k_{b}}{2}J_{b}\theta_{b}}\prod_{g}e^{\frac{k_{g}}{2}J_{g}\theta_{g}}
\prod_{i}e^{\frac{k_{r}}{2}J_{r}\Sigma^{r}_{i}\Theta^{g}_{i}\Theta^{b}_{i}+\frac{k_{b}}{2}J_{b}\Sigma^{b}_{i}\Theta^{r}_{i}\Theta^{g}_{i}+\frac{k_{g}}{2}J_{g}\Sigma^{g}_{i}\Theta^{b}_{i}\Theta^{r}_{i}}\right\}\delta(J_{r}J_{b}J_{g}-1),
\eea
\end{widetext}
where \bea \label{no}
Z_{0}&=&\frac{1}{4}(e^{\frac{N}{2}(k_{r}+k_{b}+k_{g})}+e^{\frac{N}{2}(k_{r}-k_{b}-k_{g})}+\nonumber
\\&&e^{\frac{N}{2}(-k_{r}+k_{b}-k_{g})}+e^{\frac{N}{2}(-k_{r}-k_{b}+k_{g})}) \nonumber,
\eea and $J_{c}=\pm$ subject to $J_{r}J_{b}J_{g}=1$ imposed by the
delta function. Now we are in a position that we can rewrite the
Eq.(\ref{eq15}) in terms of Ising variables regarding to the
constraint in Eq.(\ref{eq16}). Also we use the fact that
$\prod^{n}_{l=1}\sum_{J^{l}_{c}}F(J^{l}_{c})=\sum_{\{J^{l}_{c}\}^{n}_{l=1}}\prod^{n}_{l=1}F(\{J^{l}_{c}\})$.
Finally, we are left with the following expression.
\begin{widetext}                                
\bea \label{eq18}
\mathrm{Tr}[\rho^{n}_{A}]&=&\left(\frac{d_{B}}{|G|}\right)^{n-1}\frac{1}{4^{2n}Z^{n}_{0}}\sum'_{\{J^{l}_{c}\}^{n}_{l=1}}\sum'_{I_{c}}
\left(\prod_{r}\sum^{\Delta_{r}=I_{r}}_{\{\theta^{l}_{r}\}^{n}_{l=1}}e^{\frac{k_{r}}{2}\sum_{l}J^{l}_{r}\theta^{l}_{r}}\right)\left(\prod_{b}\sum^{\Delta_{b}=I_{b}}_{\{\theta^{l}_{b}\}^{n}_{l=1}}e^{\frac{k_{b}}{2}\sum_{l}J^{l}_{b}\theta^{l}_{b}}\right)\left(\prod_{g}\sum^{\Delta_{g}=I_{g}}_{\{\theta^{l}_{g}\}^{n}_{l=1}}e^{\frac{k_{g}}{2}\sum_{l}J^{l}_{g}\theta^{l}_{g}}\right)\times
\nonumber \\
&&\left(\prod_{i}{\Huge\sum^{\Delta^{c,\bar{c},\bar{\bar{c}}}_{i}=I_{c,\bar{c},\bar{\bar{c}}}}_{\{\Theta^{(l)r,b,g}_{i}\}^{n}_{l=1}}}
e^{\sum_{l}\left(\frac{k_{r}}{2}J^{l}_{r}\Sigma^{r}_{i}\Theta^{(l)g}_{i}\Theta^{(l)b}_{i}+\frac{k_{b}}{2}J^{l}_{b}\Sigma^{(l)b}_{i}\Theta^{(l)r}_{i}\Theta^{(l)g}_{i}+\frac{k_{g}}{2}J^{l}_{g}\Sigma^{g}_{i}\Theta^{(l)b}_{i}\Theta^{(l)r}_{i}\right)}\right),
 \eea
\end{widetext}
where \bea
\sum'_{\{J^{l}_{c}\}^{n}_{l=1}}&\equiv&\sum_{\{J^{l}_{r}\}^{n}_{l=1}}\sum_{\{J^{l}_{b}\}^{n}_{l=1}}\sum_{\{J^{l}_{g}\}^{n}_{l=1}}\prod^{n}_{l=1}\delta\left(J^{l}_{r}J^{l}_{b}J^{l}_{g}-1\right),
\nonumber \\ \nonumber
\sum'_{I_{c}}&\equiv&\sum_{I_{r}}\sum_{I_{b}}\sum_{I_{g}}\delta(I_{r}I_{b}I_{g}-1),
\nonumber \\ \nonumber
\Delta_{c}&=&\prod_{l}\theta^{l}_{c}~,~~\Delta^{c}_{i}=\prod_{l}\Theta^{(l)\bar{c}}_{i}\Theta^{(l)\bar{\bar{c}}}_{i}~~;~~
c=r, b, g. \eea The upper limits of sums are just implying the
constraint in Eq.(\ref{eq16}). All expressions in the above
parentheses can be restated as a partition function of a classical
Ising model. Before that, notice that the index $r$ in the typical
expression
$\left(\sum^{\Delta_{r}=I_{r}}_{\{\theta^{l}_{r}\}^{n}_{l=1}}e^{\frac{k_{r}}{2}\sum_{l}J^{l}_{r}\theta^{l}_{r}}\right)$
is mute since the sum over configuration of $\theta^{l}_{r}$'s is
done first. Other indices $b,g$ as well as $i$ are mute. Since the
$\theta^{l}$ is an Ising variable, it is possible to write it as
$\theta^{l}=\tau^{l}\tau^{l+1}$, which $\tau^{l}$'s are classical
spins. So mapping to the classical Ising model provides a tool to
have an analytical expression for entropy. The constraints
$\Delta_{r}=+1$ and $\Delta_{r}=-1$ are also satisfied by
considering the periodic or antiperiodic boundary conditions,
respectively. Thus, \bea \label{eq19}
\prod_{r}\sum^{\Delta_{r}=\pm1}_{\{\theta^{l}_{r}\}^{n}_{l=1}}e^{\frac{k_{r}}{2}\sum_{l}J^{l}_{r}\theta^{l}_{r}}=\left(\frac{1}{2}\mathcal{Z}_{n}^{p/a}(k_{r},\{J^{l}_{r}\})\right)^{\Sigma^{r}_{A}},\eea
where $\mathcal{Z}^{p/a}$ is the partition function of the classical
Ising model with periodic or antiperiodic boundary conditions. The
partition function of the Ising model can be calculated using the
transfer matrix\cite{goldenfeld} $T_{l}$ as
$\mathcal{Z}=\mathrm{Tr}[\sigma_{x}^{p/a}\prod^{n}_{l=1}T_{l}]$,
where $\sigma_{x}$ is the usual $x$-component of the Pauli matrices
and $p/a=0/1$ depending on whether we are considering the periodic
or antiperiodic boundary conditions. The corresponding partition
function is \bea \label{eq20}
\mathcal{Z}^{p/a}_{n}=2^{n}\left[\sinh^{n}(\frac{k_{r}}{2})\pm\left(\prod^{n}_{l=1}J_{r}^{l}\right)\cosh^{n}(\frac{k_{r}}{2})\right].\eea
The partition function related to the green and blue plaquettes have
similar forms. One needs only replace the index $r$ by $g$ or $b$.
The expression in the last parenthesis in Eq.(\ref{eq18}), which is
related to the collective strings, can also be mapped into a
partition function. But, the situation is rather tricky. The point
is that the terms $\Theta^{(l)g}\Theta^{(l)b}$,
$\Theta^{(l)r}\Theta^{(l)g}$ and $\Theta^{(l)b}\Theta^{(l)r}$ are
not all independent. So, it is not possible to split the exponential
function into three independent terms and then map each term to a
partition function. However, we can use the idea of mapping and
transfer matrix. In fact, in this case we use two sets of Ising
spins. Let choose $\tau$, $s$ and $\pi$ as Ising spins. Setting
$\Theta^{(l)g}\Theta^{(l)b}=\tau^{l}\tau^{l+1}$,
$\Theta^{(l)r}\Theta^{(l)g}=s^{l}s^{l+1}$ and
$\Theta^{(l)b}\Theta^{(l)r}=\pi^{l}\pi^{l+1}$ subject to the
constraint $\tau^{l}\tau^{l+1}s^{l}s^{l+1}\pi^{l}\pi^{l+1}=1$, we
hope to calculate the following partition function. Notice that
there are a 4-1 mapping between Ising spins and colored strings.
\begin{widetext} \bea
\label{eq21}\mathcal{Z}\left(k_{r}\Sigma^{r}_{i},k_{b}\Sigma^{b}_{i},k_{g}\Sigma^{g}_{i};\{J^{l}_{r}\},\{J^{l}_{b}\},\{J^{l}_{g}\}
\right)=\sum^{const.}_{\{\tau^{l},s^{l},\pi^{l}\}}e^{\sum_{l}\left(\frac{k_{r}}{2}J^{l}_{r}\Sigma^{r}_{i}\tau^{l}\tau^{l+1}+\frac{k_{b}}{2}J^{l}_{b}\Sigma^{(l)b}_{i}s^{l}s^{l+1}+
\frac{k_{g}}{2}J^{l}_{g}\Sigma^{g}_{i}\pi^{l}\pi^{l+1}\right)}. \eea
\end{widetext}
First, consider the case with constraints $\Delta^{c}_{i}=1$ for
$c=r,b,g$. The transfer matrix will be a $4\times4$ matrix as
follows:
\begin{widetext}
\bea \label{eq22}
                            T_{l}=\left(
                               \begin{array}{cccc}
                                 e^{J^{l}_{b}\textbf{b}+J^{l}_{g}\textbf{g}+J^{l}_{r}\textbf{r}} & e^{-J^{l}_{b}\textbf{b}-J^{l}_{g}\textbf{g}+J^{l}_{r}\textbf{r}} & e^{J^{l}_{b}\textbf{b}-J^{l}_{g}\textbf{g}-J^{l}_{r}\textbf{r}} & e^{-J^{l}_{b}\textbf{b}+J^{l}_{g}\textbf{g}-J^{l}_{r}\textbf{r}} \\
                                 e^{-J^{l}_{b}\textbf{b}+J^{l}_{g}\textbf{g}-J^{l}_{r}\textbf{r}} & e^{J^{l}_{b}\textbf{b}+J^{l}_{g}\textbf{g}+J^{l}_{r}\textbf{r}} & e^{-J^{l}_{b}\textbf{b}-J^{l}_{g}\textbf{g}+J^{l}_{r}\textbf{r}} & e^{J^{l}_{b}\textbf{b}-J^{l}_{g}\textbf{g}-J^{l}_{r}\textbf{r}} \\
                                 e^{J^{l}_{b}\textbf{b}-J^{l}_{g}\textbf{g}-J^{l}_{r}\textbf{r}} & e^{-J^{l}_{b}\textbf{b}+J^{l}_{g}\textbf{g}-J^{l}_{r}\textbf{r}} & e^{J^{l}_{b}\textbf{b}+J^{l}_{g}\textbf{g}+J^{l}_{r}\textbf{r}} & e^{-J^{l}_{b}\textbf{b}-J^{l}_{g}\textbf{g}+J^{l}_{r}\textbf{r}} \\
                                 e^{-J^{l}_{b}\textbf{b}-J^{l}_{g}\textbf{g}+J^{l}_{r}\textbf{r}} & e^{J^{l}_{b}\textbf{b}-J^{l}_{g}\textbf{g}-J^{l}_{r}\textbf{r}} & e^{-J^{l}_{b}\textbf{b}+J^{l}_{g}\textbf{g}-J^{l}_{r}\textbf{r}} & e^{J^{l}_{b}\textbf{b}+J^{l}_{g}\textbf{g}+J^{l}_{r}\textbf{r}} \\
                               \end{array}
                             \right),
\eea \end{widetext}

where $\textbf{b}=\frac{k_{b}}{2}\Sigma_{i}^{b}$,
$\textbf{g}=\frac{k_{g}}{2}\Sigma_{i}^{g}$ and
$\textbf{r}=\frac{k_{r}}{2}\Sigma_{i}^{r}$. The eigenvalues of the
transfer matrix can be easily calculated. Let us denote them by
$\xi^{b}_{1i}=4J^{l}_{b}\xi_{1i}$,
$\xi^{g}_{2i}=4J^{l}_{g}\xi_{2i}$, $\xi^{r}_{3i}=4J^{l}_{r}\xi_{3i}$
and $4\xi_{4i}$, where \bea \label{eq23}
\xi_{1i}=\frac{1}{2}e^{\textbf{b}}\cosh(\textbf{g}+\textbf{r})-\frac{1}{2}e^{-\textbf{b}}\cosh(\textbf{g}-\textbf{r})\nonumber
\\ \xi_{2i}=\frac{1}{2}e^{\textbf{g}}\cosh(\textbf{b}+\textbf{r})-\frac{1}{2}e^{-\textbf{g}}\cosh(\textbf{b}-\textbf{r})\nonumber
\\ \xi_{3i}=\frac{1}{2}e^{\textbf{r}}\cosh(\textbf{b}+\textbf{g})-\frac{1}{2}e^{-\textbf{r}}\cosh(\textbf{b}-\textbf{g})\nonumber
\\ \xi_{4i}=\frac{1}{2}e^{\textbf{b}}\cosh(\textbf{g}+\textbf{r})+\frac{1}{2}e^{-\textbf{b}}\cosh(\textbf{g}-\textbf{r}).\eea
Thus, regarding the constraints $\Delta^{c}_{i}=1$ for $c=r,b,g$
that are indicated by the superscript $ppp$ in the following
expression, the partition function reads as follows: \bea
\label{eq24}
\mathcal{Z}^{ppp}_{n}=4^{n}\left[J_{b}\xi^{n}_{1i}+J_{g}\xi^{n}_{2i}+J_{r}\xi^{n}_{3i}+\xi^{n}_{4i}\right],\eea
where the simplification $J_{c}=\prod^{n}_{l=1}J^{l}_{c}$ has been
used. Considering other constraints that amount to applying the
antiperiodic boundary conditions on Ising spins, the sings of
$J_{c}$ in the above expression change correspondingly. For example,
if we were to consider the
$\Delta^{b}_{i}=-\Delta^{g}_{i}=-\Delta^{r}_{i}=1$, the sings of
coefficients $J_{g}$ and $J_{r}$ become minus. Therefore, other
partition functions are as follows: \bea \label{eq25}
\mathcal{Z}^{aap}_{n}=4^{n}\left[-J_{b}\xi^{n}_{1i}+J_{g}\xi^{n}_{2i}-J_{r}\xi^{n}_{3i}+\xi^{n}_{4i}\right],\nonumber
\\ \mathcal{Z}^{paa}_{n}=4^{n}\left[-J_{b}\xi^{n}_{1i}-J_{g}\xi^{n}_{2i}+J_{r}\xi^{n}_{3i}+\xi^{n}_{4i}\right],\nonumber \\
\mathcal{Z}^{apa}_{n}=4^{n}\left[J_{b}\xi^{n}_{1i}-J_{g}\xi^{n}_{2i}-J_{r}\xi^{n}_{3i}+\xi^{n}_{4i}\right].\eea
Thus, the expression in Eq.(\ref{eq18}) is entirely given in terms
of partition functions studied here as follows:
\begin{widetext}
\bea \label{eq26}
\mathrm{Tr}[\rho^{n}_{A}]&=&\left(\frac{d_{B}}{|G|}\right)^{n-1}\frac{1}{4^{2n}Z^{n}_{0}}\frac{1}{2^{\Sigma_{A}+2m_{B}}}\sum_{\{J^{l}_{r}\}^{n}_{l=1}}\sum_{\{J^{l}_{b}\}^{n}_{l=1}}\sum_{\{J^{l}_{g}\}^{n}_{l=1}}\left[\prod^{n}_{l=1}\delta\left(J^{l}_{r}J^{l}_{b}J^{l}_{g}-1\right)\right]\times
\nonumber \\
&&\Bigg\{\left(\mathcal{Z}^{(p)}_{n}(k_{r},\{J^{l}_{r}\})\right)^{\Sigma^{r}_{A}}\left(\mathcal{Z}^{(p)}_{n}(k_{b},\{J^{l}_{b}\})\right)^{\Sigma^{b}_{A}}
\left(\mathcal{Z}^{(p)}_{n}(k_{g},\{J^{l}_{g}\})\right)^{\Sigma^{g}_{A}}\prod_{i}\mathcal{Z}^{ppp}_{n}(k_{r}\Sigma^{r}_{i},k_{b}\Sigma^{b}_{i},k_{g}\Sigma^{g}_{i};\{J^{l}_{r}\},\{J^{l}_{b}\},\{J^{l}_{g}\})
\nonumber \\&&+
\left(\mathcal{Z}^{(a)}_{n}(k_{r},\{J^{l}_{r}\})\right)^{\Sigma^{r}_{A}}\left(\mathcal{Z}^{(a)}_{n}(k_{b},\{J^{l}_{b}\})\right)^{\Sigma^{b}_{A}}
\left(\mathcal{Z}^{(p)}_{n}(k_{g},\{J^{l}_{g}\})\right)^{\Sigma^{g}_{A}}\prod_{i}\mathcal{Z}^{aap}_{n}(k_{r}\Sigma^{r}_{i},k_{b}\Sigma^{b}_{i},k_{g}\Sigma^{g}_{i};\{J^{l}_{r}\},\{J^{l}_{b}\},\{J^{l}_{g}\})\nonumber
\\&&+
\left(\mathcal{Z}^{(p)}_{n}(k_{r},\{J^{l}_{r}\})\right)^{\Sigma^{r}_{A}}\left(\mathcal{Z}^{(a)}_{n}(k_{b},\{J^{l}_{b}\})\right)^{\Sigma^{b}_{A}}
\left(\mathcal{Z}^{(a)}_{n}(k_{g},\{J^{l}_{g}\})\right)^{\Sigma^{g}_{A}}\prod_{i}\mathcal{Z}^{paa}_{n}(k_{r}\Sigma^{r}_{i},k_{b}\Sigma^{b}_{i},k_{g}\Sigma^{g}_{i};\{J^{l}_{r}\},\{J^{l}_{b}\},\{J^{l}_{g}\})
\nonumber \\&&+
\left(\mathcal{Z}^{(a)}_{n}(k_{r},\{J^{l}_{r}\})\right)^{\Sigma^{r}_{A}}\left(\mathcal{Z}^{(p)}_{n}(k_{b},\{J^{l}_{b}\})\right)^{\Sigma^{b}_{A}}
\left(\mathcal{Z}^{(a)}_{n}(k_{g},\{J^{l}_{g}\})\right)^{\Sigma^{g}_{A}}\prod_{i}\mathcal{Z}^{apa}_{n}(k_{r}\Sigma^{r}_{i},k_{b}\Sigma^{b}_{i},k_{g}\Sigma^{g}_{i};\{J^{l}_{r}\},\{J^{l}_{b}\},\{J^{l}_{g}\})\Bigg\},
\nonumber \\ \eea
\end{widetext}
where $\Sigma_{A}=\Sigma^{r}_{A}+\Sigma^{b}_{A}+\Sigma^{g}_{A}$. The
sum over $J^{l}_{c}$'s can be easily taken. Notice $J_{c}=\pm1$. Due
to the delta function, this leads to $J_{r}J_{b}J_{g}=1$. So, in
summation the $\mathrm{Z}_{2}\times \mathrm{Z}_{2}$ symmetry is
automatically held. Therefore, the factors
$J_{c}=\prod^{n}_{l=1}J^{l}_{c}$ in the above expression can be
safely dropped since the mentioned symmetry get simply exchanged the
terms between brackets. The sums give a multiplicative factor
$(\frac{1}{2}\times8)^{n}$, where the coefficient $\frac{1}{2}$
arises because of the delta function. It is convenient to introduce
the notations $x_{c}=\cosh(\frac{k_{c}}{2})$ and
$y_{c}=\sinh(\frac{k_{c}}{2})$. By inserting Eq.(\ref{eq20}),
Eq.(\ref{eq24}) and Eq.(\ref{eq25}) into Eq.(\ref{eq26}), we
eventually arrive at the following expression. \bea \label{eq27}
\mathrm{Tr}[\rho^{n}_{A}]&=&\frac{1}{4Z_{0}}
\left(\frac{d_{A}d_{B}}{Z_{0}|G|}\right)^{n-1}\times\no
\\&&\left(F^{(n)}_{1}+F^{(n)}_{2}+F^{(n)}_{3}+F^{(n)}_{4}\right), \eea
where $F$'s are functions of $x_{c}$, $y_{c}$ and $\xi$'s (see
Appendix(\ref{F})). In particular, the replica trick gives the
entanglement entropy as follows \bea \label{eq28}
S_{A}(T)&=&-\ln\left(\frac{d_{A}d_{B}}{|G|}\right)+\ln(Z_{0})\no \\
&&-\frac{1}{4Z_{0}}\left(\partial F_{1}+\partial F_{2}+\partial
F_{3}+\partial F_{4}\right), \eea where $\partial F$'s are given in
Appendix(\ref{F}). This relation is at the heart of our subsequent
discussions.
\section{limiting behavior of entanglement entropy and mutual information\label{limiting}}
Eq.(\ref{eq28}) gives all we need to explore the dependency of
entanglement entropy on temperature since apart from the first term
the remaining ones depend on temperature. First, suppose the size of
the system is finite. As the temperature goes to zero, all terms
that depend on temperature vanish and the entanglement entropy of
ground state is recovered\cite{kargarian}. The zero-temperature
entropy is $S_{A}(T\rightarrow 0)=-\ln(\frac{d_{A}d_{B}}{|G|})$.
High-temperature limit of the model corresponds to a classical model
captured through the "hard constrained limit". As the temperature
tends to infinity, the couplings $k_{c}\rightarrow \infty$ leading
to $x_{c}\sim y_{c}\sim \frac{1}{2}e^{k_{c}/2}$ and
$\xi_{1i}\sim\xi_{2i}\sim\xi_{3i}\sim\xi_{4i}\sim\frac{1}{4}e^{\textbf{b}+\textbf{r}+\textbf{g}}$.
At this limit the entanglement entropy acquires a contribution from
the bulk degrees of freedom of region $A$ as follows \bea
\label{eq29} S_{A}(T\rightarrow
\infty)=(\Sigma_{AB}+\Sigma_{A}-2m_{A})\ln2. \eea This is expected
in the sense that at high-temperature limit the thermal entropy that
scales with the volume of the region must be reached. This also
verifies that the extension of von Neumann entropy to finite
temperatures makes sense, since it contains a bulk contribution
(scaling with the volume of subsystem $A$) that corresponds to the
ordinary classical entropy. However, the above entropy, despite
being at high temperature limit, carries a constant term. This term
$2m_{A}$ depends only on the topology of the region $A$. This
exhibits even at high temperature limit the underlying system may
carry topological order. Here,the classical system is constructed by
thermalization of a pure density matrix via the high constrained
limit\cite{claudio1}. In this way precisely half of the original
topological order is preserved at the classical limit. We will refer
to this point in next section where the topological entanglement
entropy as a measure of the topological order is calculated.

Now, we turn on to take first the thermodynamic limit. In this
limit, the
entanglement entropy behaves as \bea \label{eq30} S_{A}(T)&=&-\ln\left(\frac{d_{A}d_{B}}{|G|}\right)+\frac{N}{2}(k_{r}+k_{b}+k_{g})-\ln(4)\no \\
&&-e^{-\frac{N}{2}(k_{r}+k_{b}+k_{g})}\left(\partial F_{1}+\partial
F_{2}+\partial F_{3}+\partial F_{4}\right). \no \\\eea It is simple
to check that in the thermodynamic limit $\partial F_{2},\partial
F_{3},\partial F_{4}$ tend to zero, and the above expression becomes
\bea \label{eq31}
S_{A}(T)&=&-\ln\left(\frac{d_{A}d_{B}}{|G|}\right)+\frac{N}{2}(k_{r}+k_{b}+k_{g})-\ln(4)\no
\\&-&\sum_{c}\left(e^{k_{c}/2}(x_{c}\ln x_{c}+y_{c}\ln
y_{c})-\frac{k_{c}}{2}\right)\no
\\&-&\sum_{i}e^{-\frac{k_{r}}{2}\Sigma^{r}_{p_{i}}-\frac{k_{b}}{2}\Sigma^{b}_{p_{i}}-\frac{k_{g}}{2}\Sigma^{g}_{p_{i}}}\bigg(\xi_{1i}\ln\xi_{1i}
\no
\\&+&\xi_{2i}\ln\xi_{2i}+\xi_{3i}\ln\xi_{3i}+\xi_{4i}\ln\xi_{4i}\bigg).\eea
To have a more intuitive picture of above expression, let us
consider the limit of large bipartitions which amount to
$\Sigma^{c}_{A},\Sigma^{c}_{p_{i}}\gg1$. Finally, we are left with
the following expression for the entanglement entropy \bea
\label{eq32}
S_{A}(T)_{N\rightarrow \infty}&=&(\Sigma_{AB}-2m_{A})\ln2 \no \\
&&-\sum_{c}\left[e^{k_{c}/2}(x_{c}\ln x_{c}+y_{c}\ln
y_{c})-\frac{k_{c}}{2}\right]\Sigma^{c}_{A}. \no \\ \eea If the
high-temperature limit is taken, the result coincides with
Eq.(\ref{eq29}). However, the result is not consistent with the
zero-temperature limit. That which limit is taken first would affect
the entanglement entropy as \bea \label{eq33}
\lim_{T\rightarrow0,N\rightarrow\infty}
S_{A}&=&(\Sigma_{AB}-2m_{A})\ln2, \no \\
\lim_{N\rightarrow\infty,T\rightarrow0}
S_{A}&=&(\Sigma_{AB}-2m_{A}-2m_{B}+2)\ln2. \no \\\eea Both
quantities scale with the size of the boundary, i.e. the area law
holds no matter which limit is taken first. However, the subleading
terms are different. The difference between two quantities is
$2(m_{B}-1)$, which implies that the difference depends only on the
topological properties of the regions. Like the toric code
model\cite{claudio2}, the topological contribution to the
entanglement entropy in the color code can be extracted by a single
bipartition of the model, provided the region $B$ be a
multicomponent region, i.e. $m_{B}>1$.

At zero-temperature the entanglement entropy is symmetric upon the
exchange of two subsystems $A$ and $B$, namely $S_{A}(0)=S_{B}(0)$.
But, this is no longer true at finite temperature. This is because
the entanglement entropy acquires an extensive contribution from the
bulk degrees of freedom of the region. At finite temperature the
entanglement entropy is not a measure of quantum correlations
between subsystems. The relevant quantity that drops the bulk
dependency and is symmetric between two subsystems as well is the so
called "mutual information". The mutual information grasps the total
correlations (quantum and classical) between two subsystems. In
particular, it is a linear combination of entanglement entropy as
follows \bea \label{eq34}
\mathcal{I}_{AB}(T)=\frac{1}{2}\left(S_{A}(T)+S_{B}(T)-S_{A\cup
B}(T)\right). \eea Again, the von Numann entropies for subsystems
$A$ and $B$ as well as whole system $A\cup B$ are captured by
Eq.(\ref{eq28}). Notice that $\Sigma_{A\cup B}=3N$. The derivation
of the mutual information is straitforward. However, its full
expression will be lengthy. Instead of given the full expression, we
only give the limiting behavior as for the entanglement entropy. For
finite size systems, the zero-temperature behavior coincides with
the entanglement entropy since at zero-temperature the system will
be in a pure state and the entropy $S_{A\cup B}(T)$ vanishes.
However, at the high-temperature limit, where the classical
description holds, the mutual information is symmetric between two
subsystems and scales with boundary supplemented with topological
terms as follows. \bea \label{eq35}
\mathcal{I}_{AB}=\frac{1}{2}\left(\Sigma_{AB}-2m_{A}-2m_{B}+2\right)\ln2.
\eea This quantity reveals that in the classical model precisely
half of the mutual information at zero-temperature survives. Apart
from the boundary term, the remaining terms are topological
depending on the topology of the subsystems and underlying
topological order manifests itself through them. Once again,
classical system supports half of the topological order of the
quantum system at zero-temperature. Taking first thermodynamic limit
and then zero-temperature limit don't commute with the opposite
case. While both limits scale with the boundary, the contribution of
topological terms are different. The topological contribution to the
mutual information can be filtered out by taking the difference
between two following limits: \bea \label{eq36}
\lim_{T\rightarrow0,N\rightarrow\infty}
\mathcal{I}_{AB}&=&(\Sigma_{AB}-m_{A}-m_{B}+1)\ln2, \no \\
\lim_{N\rightarrow\infty,T\rightarrow0}
\mathcal{I}_{AB}&=&(\Sigma_{AB}-2m_{A}-2m_{B}+2)\ln2. \no \\\eea The
difference \bea \label{no} \Delta
\mathcal{I}_{AB}=\lim_{T\rightarrow0,N\rightarrow\infty}
\mathcal{I}_{AB}-\lim_{N\rightarrow\infty,T\rightarrow0}
\mathcal{I}_{AB} \no \eea depends only on the topology of the
regions as $\Delta \mathcal{I}_{AB}=(m_{A}+m_{B}-1)\ln2$. The
advantage of this latter relation, in contrast to Eq.(\ref{eq33}),
is that even a simply connected bipartition of both subsystems give
rises to a nonvanishing value. So, the mutual information paves the
way in which we can define a purely topological order quantity using
a simple connected region, in contrast to linear combination over
different bipartitions\cite{kitaev2,levin}.
\section{topological entanglement entropy \label{topological}}
\subsection{finite values of $\lambda^{c}_{x}$}
Topological entanglement entropy\cite{kitaev2,levin} appears as a
subleading term of the entanglement entropy of a region with its
complement in the topological phases. It is proposed that this
subleading term can be used as a measure of topological
order\cite{toqpt}. In this section we use this measure to evaluate
the behavior of topological order versus temperature in the
topological color code. Ground state of the color code is appeared
as a superposition of two underlying string-net structures. Both of
them can be realized on the direct 2-colex lattice (unlike the toric
code where one of the loop structure is appeared on the dual
lattice). String-net structures are identified by considering the
$\sigma^{x}$ or $\sigma^{z}$ bases. The string-net structure related
to the latter bases is preserved by imposing the hard constrained
limit. But, the string-net related to the former bases can be
evaporated against the temperature. The dependency of the
topological entanglement entropy on temperature clarifies how much
robust the topological order related to the string-net structure is
against the thermal fluctuations. In this subsection we assume that
all three couplings $\lambda^{c}_{x}$ are finite.

Thermal fluctuations may be so strong that are able to break the
closed strings and leave them with end points. The broken strings as
we explained in Sec(\ref{color code}) carry excitations on their end
points. Imposing the hard constrained limit on the couplings
$\lambda^{c}_{z}\rightarrow \infty$ in the Hamiltonian in
Eq.(\ref{eq1}), the appearance of respective excitations is
restricted. So, part of topological order related to such strings is
preserved. As other couplings, $\lambda^{c}_{x}$, are finite, one
may expect that the thermal fluctuations create the star excitations
(See Fig.\ref{fig1}) at plaquettes. This means that only excitations
with charge $(e,\chi_{c})$ are created. As we will see the
accumulation of these excitations in the model destroys the
topological order partially.

As Eq.(\ref{eq28}) suggests, the entanglement entropy scales with
the boundary as well as bulk degrees of freedom. To get ride of
boundary and bulk contribution, it is convenient to use a set of
bipartitions, and then a linear combination of their entanglement
entropies unveils the topological contribution. This set of
bipartitions is shown in Fig.\ref{fig2}(b). The following linear
combination of four bipartitions gives a purely topological
contribution to the entanglement entropy. \bea \label{eq37}
S_{topo}=\lim_{R,r\rightarrow
\infty}\left(-S_{1A}+S_{2A}+S_{3A}-S_{4A}\right),\eea where each
term is given by the von Neumann entanglement entropy of the
corresponding bipartition, $R$ and $r$ are the linear size of the
subsystems as shown in Fig.\ref{fig2}(b). Large sizes of the regions
are taken to make the size of the regions much larger than the
correlation length in the topological phase. Notice that with the
above bipartitions we have the following identifications for the
number of connected and disconnected regions. \bea \label{eq38}
m_{1A}=m_{2A}=m_{3A}&=&m_{2B}=m_{3B}=m_{4B}=1, \no \\
m_{1B}&=&m_{4B}=2. \eea The following relations between number of
plaquettes related to different bipartitions also hold. \bea
\label{eq39} \Sigma_{1A}+\Sigma_{4A}=\Sigma_{2A}+\Sigma_{3A},\no
\\ \Sigma^{c}_{1A}+\Sigma^{c}_{1,1}+\Sigma^{c}_{1,2}=N, \no \\
\Sigma^{c}_{2A}+\Sigma^{c}_{2,1}=N,\no \\
\Sigma^{c}_{3A}+\Sigma^{c}_{3,1}=N,\no \\
\Sigma^{c}_{4A}+\Sigma^{c}_{4,1}=N. \eea  Notice the ambiguity to
the definition of low indices in $\Sigma^{c}_{j,i}$, where $j$
stands for one of the bipartitions in Fig.\ref{fig2}(b) and $i$ has
the same meaning as in Eq.(\ref{eq14}). The full expression of
$S_{topo}$ is very lengthy (see Appendix(\ref{Topological
Entropy})). It is rather hard to see the behavior of the topological
entanglement entropy versus temperature from this equation. To make
this quantity more clear, let us consider some limiting cases.
First, we consider the finite size systems. Zero-temperature and
high-temperature limits are as follows: \bea \label{eq40}
&&T\rightarrow0 ~(k_{c}\rightarrow0): ~~S_{topo}-S_{cc}=0, \no
\\ &&T\rightarrow\infty ~(k_{c}\rightarrow\infty): ~~S_{topo}-S_{cc}=-2\ln2,
\eea where $S_{cc}=4\ln2$ is the topological entanglement entropy of
color code at zero-temperature. In the zero-temperature limit the
model coincides with the pure ground state. However, at
high-temperature limit, as it is clear from the above relation,
precisely half of the topological entropy is removed. This result is
obtained along the hard constrained limit that has already been
taken. As we discussed in preceding section, one could expect such
result. In fact, while in the zero-temperature limit the system is
fully topological order, at the hight-temperature limit only the
string-net structures related to $Z_{c}$ plaquettes survive. Since
both string-net structures related to $X_{c}$ and $Z_{c}$ plaquettes
have the same contribution to the topological order of the ground
state, destroying one of them give rises to reduction of the
topological entanglement entropy by half.

The ground state has topological entropy $S_{cc}=\ln D^{2}$, where
$D=4$ is the so called quantum dimension of the system. By rising
the temperature, the populations of open $\gamma^{z}$ string-nets
would be favorable since the open strings anticommute with plaquette
operators $X_{c}$ that they share at odd qubits. At the very high
temperature limit,  $S_{topo}(T\rightarrow\infty)=\ln D$ implies
that each underlying string-net structure contributes $\ln D$ to the
topological entropy of the ground state.

Now, let us comment on whether the thermalization process and taking
the classical limit affect the topological sectors of the color code
model. The ground state of the color code is $4^{2\mathrm{g}}$-fold
degenerate for the systems that live on a manifold with genus
$\mathrm{g}$. The $4^{2\mathrm{g}}$ topological sectors are
identified by the eigenvalues of the non-local operators winding the
handles of the manifold. These non-local operators are product of
$\sigma^{z}$ operators along the winding closed string as
$\mathcal{S}^{c,z}_{\mu}=\prod_{i\in
\Gamma^{c}_{\mu}}\sigma_{i}^{z}$, where $\mu$ and $c$ stand for
homology and color of the respective string and $\Gamma^{c}_{\mu}$
is the support of qubits winding the handle. By a closed $c$-string
we mean a set of links that connect $c$-plaquettes. So, it commutes
with all plaquettes. Notice that for each homology class there are
three winding strings each of one color. However, because of the
interplay between color and homology only two of them are
independent. Besides, any other closed strings belonging to the same
homology class are equivalent up to a deformation.

Within each sector, the ground state is the equal superposition of
all bases obtained by any given state in the sector and applying the
plaquette operators as in Eq.(\ref{eq2}). For any ground state, the
respective totally mixed state that corresponds to the density
matrix at high-temperature limit is obtained by removing all
non-diagonal elements of pure density matrix. The value of non-local
string operators will be preserved by taking the high-temperature
limit. This latter point implies that topological sectors are not
get mixed through the thermalization of the code. In fact, for any
mixed density matrix, the expectation value of the any closed
non-winding loops that are product of plaquette operators will be
$+1$ in the hard constrained system. Classically changing the
topological sectors by flipping the spins along winding loops are
exponentially suppressed leading to the so-called broken
ergodicity\cite{claudio1}, where the phase space is divided into
topological sectors.
\begin{figure}
\begin{center}
\includegraphics[width=8cm]{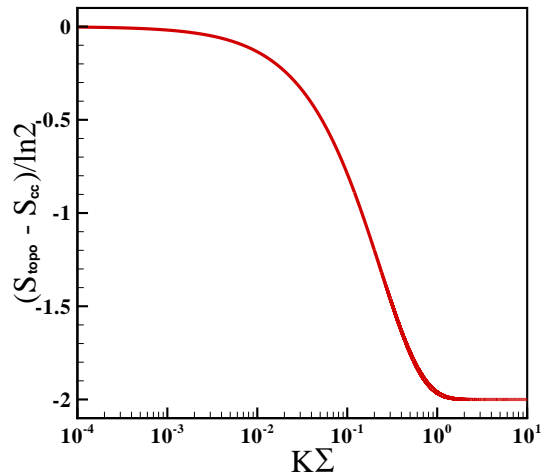}
\caption{(color online) Variation of topological entanglement
entropy in terms of coupling $K\Sigma$. The horizontal axis is in
the logarithmic scales.} \label{fig3}
\end{center}
\end{figure}
\begin{figure}
\begin{center}
\includegraphics[width=8cm]{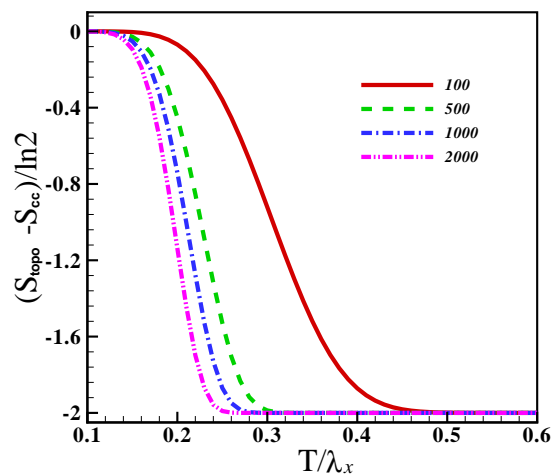}
\caption{(color online) Variation of the topological entanglement
entropy in terms of temperature. Different curves correspond to
different sizes of the inner bipartition in Fig.\ref{fig2}(b(1)). In
the limit of high temperatures all topological entropy vanishes.}

\label{fig4}
\end{center}
\end{figure}

What happens if we take the thermodynamic limit first? As the size
of the system goes to infinity, all $\Sigma^{c}_{j,i}$ tends to
infinity except one $\Sigma_{1,1}^{c}$ related to the inner region
of Fig.\ref{fig2}(b(1)). Thus, the expression in
Appendix(\ref{Topological Entropy}) gets more simplified as follows:
\bea
\label{eq41}S_{topo}&(T)&-S_{cc}(0)\rightarrow^{N\rightarrow\infty}\no
\\&&\bigg\{e^{-\frac{k_{r}}{2}\Sigma^{r}_{1,1}-\frac{k_{b}}{2}\Sigma^{b}_{1,1}-\frac{k_{g}}{2}\Sigma^{g}_{1,1}}\times(\xi^{(1)}_{11}\ln \xi^{(1)}_{11}
\no \\ &&+\xi^{(1)}_{21}\ln \xi^{(1)}_{21}+\xi^{(1)}_{31}\ln
\xi^{(1)}_{31}+\xi^{(1)}_{41}\ln \xi^{(1)}_{41})\bigg\} \no
\\&&-\frac{k_{r}}{2}\Sigma^{r}_{1,1}-\frac{k_{b}}{2}\Sigma^{b}_{1,1}-\frac{k_{g}}{2}\Sigma^{g}_{1,1}.
\eea To get more intuitive picture of above relation, we plot the
variation of $S_{topo}(T)-S_{cc}(0)$ in terms of $K\Sigma$ and
temperature in Fig.(\ref{fig3}) and Fig.(\ref{fig4}), respectively.
For simplicity we have set $K=k_{c}$ and $\Sigma=\Sigma_{1,1}^{c}$.
For finite size partition, the topological entropy drops as long as
the temperature increases. However, even at the high-temperature
limit half of topological entropy survives due to the hard
constrained limit. As it is clear in Fig.\ref{fig3} at
$K\Sigma=\frac{1}{10}$, a drop occurs at topological entropy. We use
this to estimate a characteristic temperature above which the
topological order disappears. The approximate temperature reads as
follows: \bea \label{eq42}
T_{drop}\simeq\frac{\lambda_{x}}{\ln(\sqrt{2\Sigma'})},\eea where
$\Sigma'=3\Sigma$ is the total number of plaquettes in the inner
region of Fig.\ref{fig2}(b(1)). Thus, we can conclude that the
dropping temperature depends on both coupling $\lambda_{x}$ and size
of the partition. When the size of the partition becomes large, the
dropping temperature tends to smaller values. This is transparent
from Fig.\ref{fig4} where the dropping of the topological entropy
versus temperature for different sizes of partition has been
plotted. Since we are considering the large sizes of the partition
in the very definition of the entanglement entropy, it is natural to
consider the limit of $\Sigma\gg1$ in Eq.(\ref{eq41}) that amounts
to the following relation: \bea \label{eq43}
S_{topo}&(T)&-S_{cc}(0)\rightarrow^{N\rightarrow\infty}-2\ln2,\eea
which explicitly implies that in the thermodynamic limit the
topological entropy is fragile at any non-zero temperature. This is
also expected from the dropping temperature if the bipartitions grow
proportional to each other. By this we mean that each bipartition
scales by a coefficient proportional to the size of the system. For
instance, for the bipartitions in Fig.\ref{fig2}(b(1)), we can write
$\Sigma^{c}_{A}=\gamma_{A}N$ and $\Sigma^{c}_{i}=\gamma_{i}N$, where
$0\!<\!\gamma\!<\!1$ and
$\gamma_{A}+\gamma^{c}_{1}+\gamma^{c}_{2}=1$. With this
identification, now, we can rewrite the dropping temperature in
terms of size of the system as follows:\bea \label{eq44}
T_{drop}\simeq\frac{\lambda_{x}}{\ln(\sqrt{N})},\eea which clearly
shows that as the thermodynamic limit is reached, the dropping
temperature tends to zero. In the thermodynamic limit only at zero
temperature the topological order entirely exists.
\subsection{finite values of $\lambda^{c}_{x}$ but $\lambda^{\bar{c}}_{x}$, $\lambda^{\bar{\bar{c}}}_{x}$ $\rightarrow \infty$}
Thus far, we have considered all $\lambda^{c}_{x}$'s are finite
implying that excitations are allowed in all plaquettes by rising
the temperature. As we derived in preceding subsection, at the high
temperature limit precisely half of the topological entanglement
entropy vanishes. In fact, the string-net structures related to the
three colored strings are evaporated and excitations in all
plaquettes are favorable. In the ground state, the string-net
structures contribute $2\ln D$ to the topological entropy. Now, a
question arises. How does an individual colored string impact on the
topological entropy? We give an answer to this question in what
follows. To do so, we allow for excitations to occur only on
$c$-plaquettes while excitations on other $\bar{c}$- and
$\bar{\bar{c}}$-plaquettes are restricted due to the energy cost.
This can be done by applying a similar hard constrained limit on the
couplings $\lambda^{\bar{c}}_{x}$ and $\lambda^{\bar{\bar{c}}}_{x}$
in the Hamiltonian of Eq.(\ref{eq1}). Let $c=g$, $\bar{c}=r$ and
$\bar{\bar{c}}=b$. Thus, thermal fluctuations produce only open
green strings carrying topological charges $(e,\chi_{g})$. The low
lying states of the system are
$|\psi[\gamma^{z}_{g}]\rangle=\gamma^{z}_{g}|\psi\rangle$, where
$\gamma^{z}_{g}$ is an arbitrary open green string. Within these
states, red and blue plaquettes are vortex free since
$[X_{r},\gamma^{z}_{g}]=0 $ and $[X_{b},\gamma^{z}_{g}]=0$.

Taking the limits
$\lambda^{r}_{x},\lambda^{b}_{x}\rightarrow\infty$($k_{r},k_{b}\rightarrow0$),
the topological entanglement in Eq.(\ref{eq41}) behaves as \bea
\label{eq45}
S_{topo}&(T)&-S_{cc}(0)\rightarrow^{N\rightarrow\infty}\no
\\&&\Bigg\{e^{-\frac{k_{g}}{2}\Sigma^{g}_{1,1}}\bigg[(\cosh\frac{k_{g}}{2}\Sigma^{g}_{1,1})\ln(\cosh\frac{k_{g}}{2}\Sigma^{g}_{1,1})
\no \\
&&+(\sinh\frac{k_{g}}{2}\Sigma^{g}_{1,1})\ln(\sinh\frac{k_{g}}{2}\Sigma^{g}_{1,1})
\bigg]\Bigg\}-\frac{k_{g}}{2}\Sigma^{g}_{1,1}.\no \\\eea As we are
eventually interested in large sizes of the bipartitions, i.e.
$\Sigma^{g}_{1,1}\gg1$, any non-zero temperature subsides the
topological entanglement as follows\bea \label{eq46}
S_{topo}&(T)&-S_{cc}(0)\rightarrow^{N\rightarrow\infty}-\ln2. \eea
For finite size partitions and in the high-temperature limit, as
long as $\lambda^{r,b}_{x}/T\gg1$, the same reduction in the
topological entropy occurs. It is obvious from the relation in
Eq.(\ref{eq46}) that with disruption of one type of strings, say
green, topological order vanishes partially. This is not same as in
Eq.(\ref{eq43}), where all closed colored strings are allowed to
evaporate.

\begin{figure}
\begin{center}
\includegraphics[width=8cm]{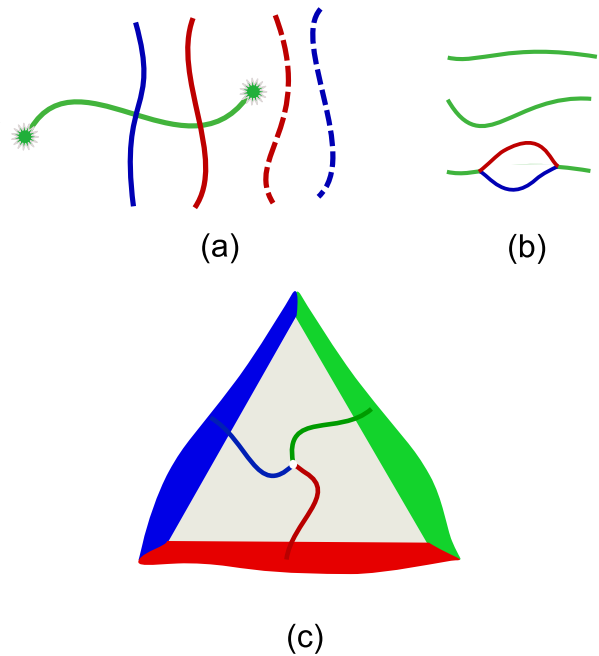}
\caption{(color online) (a) An open green (light gray) string
carrying star excitations on its ends. The winding blue(dark gray)
and red(gray) wavy lines on either sides of the excitation pick up
different values. (b) From top to bottom: a colored string, here
green, can be either deform or split into red and blue strings. (c)
A manifestation of a 2-colex with boundary. Here, we consider a
triangular code with three borders: blue(dark gray), red(gray) and
green(light gray). A basis for this code is a string-net in which
each string has an end point at the border with same color. }
\label{fig5}
\end{center}
\end{figure}

In order to understand the dependency of the topological entropy to
the colored strings, let us consider the non-local winding string
operators as order parameters. They are products of pauli operators
along paths winding the torus as
$\mathcal{S}^{c,x}_{\mu}=\prod_{i\in
\Gamma^{c}_{\mu}}\sigma_{i}^{x}$. Consider the ground state of the
system in the $\sigma_{i}^{x}$ bases. The excited states are then
$|\psi[\gamma^{z}_{g}]\rangle=\gamma^{z}_{g}|\psi\rangle$. If a
non-local string and an open green string $\gamma^{z}_{g}$ cross
each other odd times and have different colors then
$\{\mathcal{S}^{c,x}_{\mu},\gamma^{z}_{g}\}=0$. Otherwise they
commute. At the ground state level, the expectation values of
non-local strings are nonzero and independent of any deformation of
the strings, i.e.
$\Upsilon^{c}(0)=\langle\psi|\mathcal{S}^{c,x}_{\mu}|\psi\rangle=+1(-1)$,
depending on the topological sector we are analyzing it. But, at
finite temperature the expectation value is replaced by thermal
average as
$\Upsilon^{c}(T)=\frac{1}{N_{\Gamma}}\sum_{\{\Gamma\}}\langle\mathcal{S}^{c,x}_{\mu}\rangle_{T}$.
The excitations carrying by open green strings are deconfined and
the expectation value of two non-local red (blue) strings on
opposite sides of excitations are different. For instance consider
an open green string carrying star excitations on its ends as shown
in Fig.\ref{fig5}(a). The winding blue and red strings cross the
green string either odd times (solid wavy lines) or nothing (dashed
wavy lines). The solid lines anticommute with green string while
dashed lines commute. In the expectation value we must take average
over all possible cases. So at finite temperature, the emerging
excitations destroy the non-local order parameter
$\Upsilon^{r,b}(T)\simeq0$. However, the expectation value of a
non-local green string remains finite since it commutes with
defects. These lead to destroying the string-net structure of the
ground state through the thermalization. The thermal states still
contain closed green strings since they commute with defects, i.e.
$[\Lambda^{x}_{g},\gamma^{z}_{g}]=0$, where $\Lambda^{x}_{g}$ is an
arbitrary product of $\sigma^{x}$ operators living on a closed green
string. Notice that closed green strings are product of red and blue
plaquettes  $X_{r}$ and $X_{b}$ for which the expectation values
with respect to thermal states are $+1$, i.e. they are stabilized by
red and blue plaquettes. Thus, one may expect that the topological
order in $\sigma^{x}$ bases is partially preserved. This is just the
message of the Eq.(\ref{eq46}) that topological entanglement entropy
can be reexpressed into $S_{topo}=\ln D+\frac{1}{2}\ln D$, where
$\ln D$ is due to the string-net structure related to the
$\sigma^{z}$ bases and $\frac{1}{2}\ln D$ is ascribed to one type of
closed colored strings (here green) in $\sigma^{x}$ bases survived
even at finite temperature.

It is tempting to infer that each colored string contribute
$\frac{1}{2}\ln D$ to the topological entropy. However, the string
structure of the topological order in the ground state of the color
code is subject to the $\mathbf{Z_{2}}\times \mathbf{Z_{2}}$ gauge
symmetry. This is a property of the color code that each colored
string is a $\mathbf{Z_{2}}$ gauge degrees of freedom, but all
strings form a $\mathbf{Z_{2}}\times \mathbf{Z_{2}}$ gauge structure
that is rooted in the string-net structure of the model. Nature of
the topological order in the ground state comes from the fact that
the ground state is invariant not only under deformation of strings
but also the splitting of a $c$-string into $\bar{c}$- and
$\bar{\bar{c}}$-strings as shown in Fig.\ref{fig5}(b). The latter
point about splitting corresponds to the structure of the gauge
group of the color code as we explained in Sec.(\ref{color code}).

The relation between topological entropy and colored strings can be
further understood if we soften more couplings. For example let
$\lambda^{g}_{x}$, $\lambda^{r}_{x}$ be finite while
$\lambda^{b}_{x}/T\gg1$. Now, the creation of defects in green and
red plaquettes becomes favorable. Although the blue plaquettes
remain immune against the thermal fluctuations, the average value of
all colored winding strings vanishes, since they anticommute with
green or red defects. Thus, all topological order vanishes. This can
also be seen from the Eq.(\ref{eq41}) by letting
$\lambda^{b}_{x}/T\gg1$($k_{b}\rightarrow0$). In the limit of large
bipartitions, all topological order vanishes as follows \bea
\label{eq47}
S_{topo}&(T)&-S_{cc}(0)\rightarrow^{N\rightarrow\infty}-2\ln2. \eea
Once again when two colored strings are allowed to evaporate, the
topological order in the system subsides. Putting all things
together, we can conclude that the topological entanglement entropy
related to the string-net structure in the $\sigma^{x}$ bases
receives contributions from the colored strings. In fact, each
colored string contributes $\frac{1}{2}\ln D$ to the topological
entropy, but the symmetry of the color code model gives rise to the
total contribution $2\times \frac{1}{2}\ln D$.

\section{open boundary conditions: planar colore codes \label{triangular}}
Up to this point, we have only considered the system embedded in a
closed surface such as torus, i.e. with periodic boundary
conditions. However, the physical system must have boundary in which
we can confine to a certain piece of space. It is simple to obtain
an open surface from a closed one by removing some plaquettes of the
2-colex. Such lattices are called planar color code. For example,
consider a 2-colex embedded on a sphere. In this case the lattice
encodes zero qubits, i.e. the ground state subspace is spanned by a
single vector. If we remove a single qubit and its three neighboring
links and plaquettes, the obtained 2-colex will encode a single
qubit. In fact, three missed plaquettes form three colored borders
for the lattice, and only strings with the same color of border will
have end points on that border. An example of such bordered lattice
is shown in Fig.\ref{fig5}(c).

The most important class of the planar color code is the so called
triangular color code\cite{hector1}. The triangular code has three
colored borders each of one color (See Fig.\ref{fig5}(c)). The
logical operators are given by a string-net and its deformation in
which make a two-dimensional algebra, since the code encodes only a
single qubit. Such string-net is crucial in implementation of the
Clifford group. The stabilizer group of the code is again given by a
set of plaquettes. However, all plaquettes are independent because
of boundary, which affect the number of collective operators needed
for evaluating of entanglement entropy. Although a 2-colex on closed
and open surfaces may represent different properties for the
encoding and implementation of quantum information processing, they
represent the same entanglement properties as we will see what
follows. This is due to the fact that both structures have same
symmetry and topological order comes from the string-net structure
of the model.

The group element $\tilde{h}$ can be produced only in one way that
is the product of some plaquettes. This amounts to consider only one
of the terms in Eq.(\ref{eq7}) and Eq.(\ref{eq9}). This leads to the
following simple relation for $\eta_{T}(\tilde{h})$. \bea
\label{eq48}\eta_{T}(\tilde{h})=e^{-k_{g}n_{g}-k_{b}n_{b}-k_{r}n_{r}}.
\eea We can follow the method in Sec(\ref{entanglement entropy}) to
map the contribution of plaquettes and collective strings in the
group element into Ising variables. However, we should take care
about the collective strings. Consider two disjoint regions $B_{1}$
and $B_{2}$ and connected region $A$ as in Fig.\ref{fig2}(a), but,
on an open surface. Each disjoint region is surrounded by three
colored strings in which only two of them are independent. Observe
that two colored strings, say red, surrounding regions $B_{1}$ and
$B_{2}$ are independent, in contrast to the closed surface. By this
we mean that if $\gamma^{r}_{i}$ be a collective string around the
disjoint region $B_{i}$, the $\prod_{i}\gamma^{r}_{i}$ is not the
product of blue and green plaquettes of region $A$. Implication of
this point arises in the cardinalities of subgroups $G_{A}$ and
$G_{B}$.

Since the open boundary conditions establish only one construction
for an group element $\tilde{h}$ in terms of plaquette operators,
the preceding arguments leading to the expression in Eq.(\ref{eq26})
give only the first term as follows:
\begin{widetext}
\bea \label{eq49}
\mathrm{Tr}[\rho^{n}_{A}]&=&\left(\frac{d_{B}}{|G|}\right)^{n-1}\frac{1}{Z^{n}_{1}}\frac{1}{2^{\Sigma_{A}+2m_{B}}}\sum_{\{J^{l}_{r}\}^{n}_{l=1}}\sum_{\{J^{l}_{b}\}^{n}_{l=1}}\sum_{\{J^{l}_{g}\}^{n}_{l=1}}\left[\prod^{n}_{l=1}\delta\left(J^{l}_{r}J^{l}_{b}J^{l}_{g}-1\right)\right]\times
\nonumber \\
&&\Bigg\{\left(\mathcal{Z}^{(p)}_{n}(k_{r},\{J^{l}_{r}\})\right)^{\Sigma^{r}_{A}}\left(\mathcal{Z}^{(p)}_{n}(k_{b},\{J^{l}_{b}\})\right)^{\Sigma^{b}_{A}}
\left(\mathcal{Z}^{(p)}_{n}(k_{g},\{J^{l}_{g}\})\right)^{\Sigma^{g}_{A}}\prod_{i}\mathcal{Z}^{ppp}_{n}(k_{r}\Sigma^{r}_{i},k_{b}\Sigma^{b}_{i},k_{g}\Sigma^{g}_{i};\{J^{l}_{r}\},\{J^{l}_{b}\},\{J^{l}_{g}\})\Bigg\},
\nonumber \\\eea \end{widetext} where
$Z_{1}=e^{\frac{N}{2}(k_{g}+k_{b}+k_{r})}$. In particular, the
replica gives the entanglement entropy as \bea \label{eq50}
S_{A}(T)&=&-\ln\left(\frac{d_{A}d_{B}}{|G|}\right)+\ln(Z_{1})-\frac{1}{Z_{1}}\partial
F_{1}. \eea The expression $\partial F_{1}$ is given in
Appendix(\ref{F}).

For finite size systems, taking the zero-temperature limit gives the
entanglement entropy of a single bipartition, $m_{A}=m_{B}=1$, in
the ground state of the triangular color code, i.e.
$S^{t}_{A}(0)=(\Sigma_{AB}-2)\ln2$, where the upper index $t$
denotes the triangular code. If we were to take first the
thermodynamic limit and then the zero- temperature limit, the limits
don't commute with each other. The difference between two limits
depends only on the topology of the regions as $\Delta
S^{t}=\lim_{T\rightarrow0,N\rightarrow\infty}S_{A}-\lim_{N\rightarrow\infty,T\rightarrow0}S^{t}_{A}=2m_{B}\ln2$,
which implies that the topological contribution to the entanglement
entropy can be extracted even by a single bipartition.

Topological entanglement entropy is given by Eq.(\ref{eq37}) and the
respective bipartitions in Fig.\ref{fig2}(b). We should take into
account the restriction imposed by boundary. The cardinalities of
bipartitions in Fig.\ref{fig2}(b(1)) are $d_{A}=2^{\Sigma_{1A}+2}$
and $d_{B}=2^{\Sigma_{1B}+2}$. The corresponding quantities for
bipartitions in Fig.\ref{fig2}(b(2)) and Fig.\ref{fig2}(b(3)) are
$d_{A}=2^{\Sigma_{2,3A}}$ and $d_{B}=2^{\Sigma_{2,3B}+2}$. Last
bipartition in Fig.\ref{fig2}(b(4)) yields the cardinalities
$d_{A}=2^{\Sigma_{4A}}$ and $d_{B}=2^{\Sigma_{4B}+4}$. This leads to
the topological entropy of triangular color code at zero
temperature\cite{kargarian} as $S^{t}_{cc}=4\ln2$, which is similar
with that of the periodic boundary conditions.

Finite temperature topological entropy behaves as what we derived
for closed surface. For finite size systems and at high-temperature
limit precisely half of the topological entanglement is preserved as
$S^{t}_{topo}(T\rightarrow\infty)-S^{t}_{cc}=-2\ln2$, which is a
consequence of destroying of string-net structure in the
$\sigma^{x}$ or $\sigma^{z}$ bases. If we take first the
thermodynamic limit, any non-zero temperature subsides the
topological entropy to half of its value at zero temperature.
\section{conclusions \label{conclusion}}
In this work the topological order in a class of two-dimensional
topological stabilizer codes, the so called color code, at finite
temperature has been addressed. Both closed and open surfaces, where
the lattice embedded on, were considered. This is because
quasiparticle excitations are different for each embedding. The
stabilizer structure of the color code comes from the plaquette
operators which can be of $X$ and $Z$ Pauli operators. The
plaquettes are labeled by their color and type. The stabilizer group
is adjusted into a many-body Hamiltonian in which the coding space
is its ground state subspace. The Hamiltonian is supplemented with
energy scales such as $\lambda^{c}_{z}$'s and $\lambda^{c}_{x}$'s.
The ground state is a string-net condensate. Closed string-nets are
collections of colored strings in which no end points are left.
Topological order in the ground state of the model arises from the
coherent superposition of two string-net structures that can be
visualized by adopting the string-nets in $\sigma^{z}$ and
$\sigma^{x}$ bases. This topological order can be characterized by
using the topological entanglement entropy that is $2\ln D$ where
$D=4$.

Considering the limit $\lambda^{c}_{z}\rightarrow\infty$, the exact
solution of the model at finite temperature becomes possible. As the
temperature is increased, for finite size systems, the topological
entropy reduced to $\ln D$. This implies that by thermalization of
one of the string-net structures the respective topological order is
destroyed. Both string-net structures have equal contribution to
topological entropy.

The temperature at which the topological entropy is dropped is a
function of both coupling $\lambda_{x}$ and size of system as in
Eq.(\ref{eq44}). This relation can be used to distinguish a length
scale for separated defects. It can be recasted into
$Ne^{-2\lambda_{x}/T_{drop}}\simeq1$. This means that topological
order is destroyed when density of defects in the model becomes of
order unity as the temperature is increased. Similar behavior
observed in the toric code\cite{claudio2}. This interpretation for
density of defects allows us to define the respective length scale
as $\zeta_{x}=e^{\lambda_{x}/T}$. For temperatures well below the
dropping temperature $T\ll T_{drop}$, the characteristic length
scale is large implying the density of defects
$Ne^{-2\lambda_{x}/T}$ is much less than unity. In this case the
thermal defects are not able to shave out the string-net structure
of the model and topological order is preserved. As the temperature
increases, the density of defects becomes of order unity that
destroy the topological order in the system. In the thermodynamic
limit and at finite temperature the distance between defects is much
less than the system size. Thus at this limit any nonzero
temperature can destroy topological order as in Eq.(\ref{eq43}). The
robustness of the string-net structure in the hard constrained limit
can be understood via such identification for distance between
defects. In the limit $\lambda_{z}\rightarrow\infty$ the distance
between respective defects $\zeta_{z}=e^{\lambda_{z}/T}$ is
infinity. So, the respective topological order is immune even in the
high temperature limit. This is the reason why in the high
constrained system half of the topological order is preserved.

Topological order in the color code arises from coherent
superposition of string-net structures in $\sigma^{x}$ and
$\sigma^{z}$ bases. Both structures are similar having the same
contribution in the topological entropy. By imposing extra
conditions on the couplings $\lambda^{c}_{x}$, we can examine the
contribution of colors and underlying symmetry on the topological
entropy. If we let for defects to occur only in $c$-plaquettes, i.e.
$\lambda^{\bar{c}}_{x},
\lambda^{\bar{\bar{c}}}_{x}\rightarrow\infty$, the thermal states
still carry topological order in the $\sigma^{x}$ bases even at
thermodynamic limit as in Eq.(\ref{eq46}). However, if we soften the
latter condition and let for defects to occur in both $c$- and
$\bar{c}$-plaquettes, i.e. only let
$\lambda^{\bar{\bar{c}}}_{x}\rightarrow\infty$, as Eq.(\ref{eq47})
suggests topological order is entirely destroyed. These observations
reveal that both colors and symmetry determine the topological
entropy. In each loop structure, either in $\sigma^{x}$ or
$\sigma^{z}$ bases, string-nets are collections of closed colored
strings. Each colored closed string contributes $\frac{1}{2}\ln D$
to the topological entropy. Considering the gauge symmetry of the
model, the total contributions of closed colored strings in the
topological entropy will be $\ln D$. In the ground state both
string-net structures in $\sigma^{x}$ and $\sigma^{z}$ bases
contribute equally yielding the topological entropy as $2\ln D$.

All above results also hold for the case of open boundary
conditions. For this case we considered the triangular color code
with borders. Although lattices embedded in closed or open surfaces
present different properties from the quantum information
perspective, they have similar topological order. This is because
they have similar gauge symmetry group and consequently same
underlying string structure that is reflected in topological
entropy.

The topological order in both toric code and color code models in
two-dimensional space is fragile against thermal fluctuations in the
sense that their underlying structures are one-dimensional objects,
i.e. the strings. This fragility limits their capabilities as
self-stability codes. The deconfinement of open strings carrying
excitations can be restricted by coupling the defects to bosonic
fields\cite{hamma2}. Toric code model in three-dimension have a
membrane structure that are robust against thermal
noises\cite{claudio3}. The color code model can also be generalized
to $D$-dimensional space with branyon and brane-net structures. It
is instructive to generalize the approach presented here to color
codes based on the D-colexes\cite{miguel1,miguel3} in the sense that
they have different underlying objects as physical
mechanism for topological order.\\

\section{acknowledgement}
I would like to thank A. Langari for fruitful discussions and
comments. I would also like to acknowledge M. A. Martin-Delgado and
H. Bombin for their useful comments. This work was supported in part
by the Center of Excellence in Complex Systems and Condensed Matter
(www.cscm.ir).\\\\

\section{Appendix \label{Appendix}}

\subsection{F's expressions and their derivatives\label{F}}
$F^{(n)}_{j}$'s for $j=1,2,3,4$ stand for four terms between the two
brackets in Eq.(\ref{eq26}). By using the partition functions
related to plaquettes and strings that are given in Eq.(\ref{eq20}),
Eq.(\ref{eq24}) and Eq.(\ref{eq25}), the following relations for
$F^{(n)}_{j}$'s are obtained.
\begin{widetext}
\bea \label{app1}
F^{(n)}_{1}=(x^{n}_{r}+y^{n}_{r})^{\Sigma^{r}_{A}}(x^{n}_{b}+y^{n}_{b})^{\Sigma^{b}_{A}}(x^{n}_{g}+y^{n}_{g})^{\Sigma^{g}_{A}}\prod_{i}(\xi^{n}_{1i}+\xi^{n}_{2i}+\xi^{n}_{3i}+\xi^{n}_{4i}) \no \\
F^{(n)}_{2}=(x^{n}_{r}-y^{n}_{r})^{\Sigma^{r}_{A}}(x^{n}_{b}-y^{n}_{b})^{\Sigma^{b}_{A}}(x^{n}_{g}+y^{n}_{g})^{\Sigma^{g}_{A}}\prod_{i}(-\xi^{n}_{1i}+\xi^{n}_{2i}-\xi^{n}_{3i}+\xi^{n}_{4i})\no \\
F^{(n)}_{3}=(x^{n}_{r}+y^{n}_{r})^{\Sigma^{r}_{A}}(x^{n}_{b}-y^{n}_{b})^{\Sigma^{b}_{A}}(x^{n}_{g}-y^{n}_{g})^{\Sigma^{g}_{A}}\prod_{i}(-\xi^{n}_{1i}-\xi^{n}_{2i}+\xi^{n}_{3i}+\xi^{n}_{4i}) \no \\
F^{(n)}_{4}=(x^{n}_{r}-y^{n}_{r})^{\Sigma^{r}_{A}}(x^{n}_{b}+y^{n}_{b})^{\Sigma^{b}_{A}}(x^{n}_{g}-y^{n}_{g})^{\Sigma^{g}_{A}}\prod_{i}(\xi^{n}_{1i}-\xi^{n}_{2i}-\xi^{n}_{3i}+\xi^{n}_{4i})
 \eea
\end{widetext}
The case $n=1$ and their derivatives are needed for evaluation of
entanglement entropy. So, the above expressions get more simplified
as follows:
\bea \label{no} &&F^{(1)}_{1}=e^{\frac{N}{2}(k_{r}+k_{b}+k_{g})},~~
F^{(1)}_{2}=e^{\frac{N}{2}(-k_{r}-k_{b}+k_{g})},~~ \no \\
&&F^{(1)}_{3}=e^{\frac{N}{2}(k_{r}-k_{b}-k_{g})},~~
F^{(1)}_{4}=e^{\frac{N}{2}(-k_{r}+k_{b}-k_{g})}. \no \eea
Trivially
$Z_{0}=\frac{1}{4}(F^{(1)}_{1}+F^{(1)}_{2}+F^{(1)}_{3}+F^{(1)}_{3})$,
and their derivatives at $n=1$ are
\begin{widetext}
\bea \label{app2}
\partial F_{1}=(\partial F^{(n)}_{1}/\partial n)_{n \rightarrow
1}&=&F^{(1)}_{1}\Bigg[\Sigma^{r}_{A}e^{\frac{-k_{r}}{2}}(x_{r}\ln
x_{r}+y_{r}\ln y_{r})+\Sigma^{b}_{A}e^{\frac{-k_{b}}{2}}(x_{b}\ln
x_{b}+y_{b}\ln y_{b})+\Sigma^{g}_{A}e^{\frac{-k_{g}}{2}}(x_{g}\ln
x_{g}+y_{g}\ln y_{g})\no \\
&&+
\sum_{i}e^{-\frac{k_{r}}{2}\Sigma^{r}_{i}-\frac{k_{b}}{2}\Sigma^{b}_{i}-\frac{k_{g}}{2}\Sigma^{g}_{i}}(\xi_{1i}\ln
\xi_{1i}+\xi_{2i}\ln \xi_{2i}+\xi_{3i}\ln \xi_{3i}+\xi_{4i}\ln
\xi_{4i})\Bigg], \no \\
\partial F_{2}=(\partial F^{(n)}_{2}/\partial n)_{n \rightarrow
1}&=&F^{(2)}_{2}\Bigg[\Sigma^{r}_{A}e^{\frac{k_{r}}{2}}(x_{r}\ln
x_{r}-y_{r}\ln y_{r})+\Sigma^{b}_{A}e^{\frac{k_{b}}{2}}(x_{b}\ln
x_{b}-y_{b}\ln y_{b})+\Sigma^{g}_{A}e^{\frac{-k_{g}}{2}}(x_{g}\ln
x_{g}+y_{g}\ln y_{g})\no \\
&&+
\sum_{i}e^{\frac{k_{r}}{2}\Sigma^{r}_{i}+\frac{k_{b}}{2}\Sigma^{b}_{i}-\frac{k_{g}}{2}\Sigma^{g}_{i}}(-\xi_{1i}\ln
\xi_{1i}+\xi_{2i}\ln \xi_{2i}-\xi_{3i}\ln \xi_{3i}+\xi_{4i}\ln
\xi_{4i})\Bigg], \no \\
\partial F_{3}=(\partial F^{(n)}_{3}/\partial n)_{n \rightarrow
1}&=&F^{(3)}_{3}\Bigg[\Sigma^{r}_{A}e^{\frac{-k_{r}}{2}}(x_{r}\ln
x_{r}+y_{r}\ln y_{r})+\Sigma^{b}_{A}e^{\frac{k_{b}}{2}}(x_{b}\ln
x_{b}-y_{b}\ln y_{b})+\Sigma^{g}_{A}e^{\frac{k_{g}}{2}}(x_{g}\ln
x_{g}-y_{g}\ln y_{g})\no \\
&&+
\sum_{i}e^{-\frac{k_{r}}{2}\Sigma^{r}_{i}+\frac{k_{b}}{2}\Sigma^{b}_{i}+\frac{k_{g}}{2}\Sigma^{g}_{i}}(-\xi_{1i}\ln
\xi_{1i}-\xi_{2i}\ln \xi_{2i}+\xi_{3i}\ln \xi_{3i}+\xi_{4i}\ln
\xi_{4i})\Bigg], \no \\
\partial F_{4}=(\partial F^{(n)}_{4}/\partial n)_{n \rightarrow
1}&=&F^{(4)}_{4}\Bigg[\Sigma^{r}_{A}e^{\frac{k_{r}}{2}}(x_{r}\ln
x_{r}-y_{r}\ln y_{r})+\Sigma^{b}_{A}e^{\frac{-k_{b}}{2}}(x_{b}\ln
x_{b}+y_{b}\ln y_{b})+\Sigma^{g}_{A}e^{\frac{k_{g}}{2}}(x_{g}\ln
x_{g}-y_{g}\ln y_{g})\no \\
&&+
\sum_{i}e^{\frac{k_{r}}{2}\Sigma^{r}_{i}-\frac{k_{b}}{2}\Sigma^{b}_{i}+\frac{k_{g}}{2}\Sigma^{g}_{i}}(\xi_{1i}\ln
\xi_{1i}-\xi_{2i}\ln \xi_{2i}-\xi_{3i}\ln \xi_{3i}+\xi_{4i}\ln
\xi_{4i})\Bigg], \no \\
\eea
\end{widetext}

\subsection{Topological Entropy \label{Topological Entropy}}

Applying Eq.(\ref{eq28}) about each bipartition of
Fig.(\ref{fig2}b), and inserting into Eq.(\ref{eq37}), the
topological entanglement entropy then reads as follows:
\begin{widetext}
\bea \label{app3} S_{topo}(T)&-&S_{cc}(0)=  \no \\
&&\frac{1}{4Z_{0}}\sum^{2}_{i=1}\bigg\{e^{\frac{k_{r}}{2}(N-\Sigma^{r}_{1,i})+\frac{k_{b}}{2}(N-\Sigma^{b}_{1,i})+\frac{k_{g}}{2}(N-\Sigma^{g}_{1,i})}
(\xi^{(1)}_{1i}\ln \xi^{(1)}_{1i}+\xi^{(1)}_{2i}\ln
\xi^{(1)}_{2i}+\xi^{(1)}_{3i}\ln \xi^{(1)}_{3i}+\xi^{(1)}_{4i}\ln
\xi^{(1)}_{4i}) \no
\\&&+e^{-\frac{k_{r}}{2}(N-\Sigma^{r}_{1,i})-\frac{k_{b}}{2}(N-\Sigma^{b}_{1,i})+\frac{k_{g}}{2}(N-\Sigma^{g}_{1,i})}
(-\xi^{(1)}_{1i}\ln \xi^{(1)}_{1i}+\xi^{(1)}_{2i}\ln
\xi^{(1)}_{2i}-\xi^{(1)}_{3i}\ln \xi^{(1)}_{3i}+\xi^{(1)}_{4i}\ln
\xi^{(1)}_{4i}) \no \\
&&+e^{\frac{k_{r}}{2}(N-\Sigma^{r}_{1,i})-\frac{k_{b}}{2}(N-\Sigma^{b}_{1,i})-\frac{k_{g}}{2}(N-\Sigma^{g}_{1,i})}
(-\xi^{(1)}_{1i}\ln \xi^{(1)}_{1i}-\xi^{(1)}_{2i}\ln
\xi^{(1)}_{2i}+\xi^{(1)}_{3i}\ln \xi^{(1)}_{3i}+\xi^{(1)}_{4i}\ln
\xi^{(1)}_{4i}) \no \\
&&+e^{-\frac{k_{r}}{2}(N-\Sigma^{r}_{1,i})+\frac{k_{b}}{2}(N-\Sigma^{b}_{1,i})-\frac{k_{g}}{2}(N-\Sigma^{g}_{1,i})}
(\xi^{(1)}_{1i}\ln \xi^{(1)}_{1i}-\xi^{(1)}_{2i}\ln
\xi^{(1)}_{2i}-\xi^{(1)}_{3i}\ln \xi^{(1)}_{3i}+\xi^{(1)}_{4i}\ln
\xi^{(1)}_{4i})\bigg\} \no \\
&&-\frac{1}{4Z_{0}}
\bigg\{e^{\frac{k_{r}}{2}(N-\Sigma^{r}_{2,1})+\frac{k_{b}}{2}(N-\Sigma^{b}_{2,1})+\frac{k_{g}}{2}(N-\Sigma^{g}_{2,1})}
(\xi^{(2)}_{1}\ln \xi^{(2)}_{1}+\xi^{(2)}_{2}\ln
\xi^{(2)}_{2}+\xi^{(2)}_{3}\ln \xi^{(2)}_{3}+\xi^{(2)}_{4}\ln
\xi^{(2)}_{4}) \no
\\&&+e^{-\frac{k_{r}}{2}(N-\Sigma^{r}_{2,1})-\frac{k_{b}}{2}(N-\Sigma^{b}_{2,1})+\frac{k_{g}}{2}(N-\Sigma^{g}_{2,1})}
(-\xi^{(2)}_{1}\ln \xi^{(2)}_{1}+\xi^{(2)}_{2}\ln
\xi^{(2)}_{2}-\xi^{(2)}_{3}\ln \xi^{(2)}_{3}+\xi^{(2)}_{4}\ln
\xi^{(2)}_{4}) \no \\
&&+e^{\frac{k_{r}}{2}(N-\Sigma^{r}_{2,1})-\frac{k_{b}}{2}(N-\Sigma^{b}_{2,1})-\frac{k_{g}}{2}(N-\Sigma^{g}_{2,1})}
(-\xi^{(2)}_{1}\ln \xi^{(2)}_{1}-\xi^{(2)}_{2}\ln
\xi^{(2)}_{2}+\xi^{(2)}_{3}\ln \xi^{(2)}_{3}+\xi^{(2)}_{4}\ln
\xi^{(2)}_{4}) \no \\
&&+e^{-\frac{k_{r}}{2}(N-\Sigma^{r}_{2,1})+\frac{k_{b}}{2}(N-\Sigma^{b}_{2,1})-\frac{k_{g}}{2}(N-\Sigma^{g}_{2,1})}
(\xi^{(2)}_{1}\ln \xi^{(2)}_{1}-\xi^{(2)}_{2}\ln
\xi^{(2)}_{2}-\xi^{(2)}_{3}\ln \xi^{(2)}_{3}+\xi^{(2)}_{4}\ln
\xi^{(2)}_{4})\bigg\} \no \\
&&-\frac{1}{4Z_{0}}
\bigg\{e^{\frac{k_{r}}{2}(N-\Sigma^{r}_{3,1})+\frac{k_{b}}{2}(N-\Sigma^{b}_{3,1})+\frac{k_{g}}{2}(N-\Sigma^{g}_{3,1})}
(\xi^{(3)}_{1}\ln \xi^{(3)}_{1}+\xi^{(3)}_{2}\ln
\xi^{(3)}_{2}+\xi^{(3)}_{3}\ln \xi^{(3)}_{3}+\xi^{(3)}_{4}\ln
\xi^{(3)}_{4}) \no
\\&&+e^{-\frac{k_{r}}{2}(N-\Sigma^{r}_{3,1})-\frac{k_{b}}{2}(N-\Sigma^{b}_{3,1})+\frac{k_{g}}{2}(N-\Sigma^{g}_{3,1})}
(-\xi^{(3)}_{1}\ln \xi^{(3)}_{1}+\xi^{(3)}_{2}\ln
\xi^{(3)}_{2}-\xi^{(3)}_{3}\ln \xi^{(3)}_{3}+\xi^{(3)}_{4}\ln
\xi^{(3)}_{4}) \no \\
&&+e^{\frac{k_{r}}{2}(N-\Sigma^{r}_{3,1})-\frac{k_{b}}{2}(N-\Sigma^{b}_{3,1})-\frac{k_{g}}{2}(N-\Sigma^{g}_{3,1})}
(-\xi^{(3)}_{1}\ln \xi^{(3)}_{1}-\xi^{(3)}_{2}\ln
\xi^{(3)}_{2}+\xi^{(3)}_{3}\ln \xi^{(3)}_{3}+\xi^{(3)}_{4}\ln
\xi^{(3)}_{4}) \no \\
&&+e^{-\frac{k_{r}}{2}(N-\Sigma^{r}_{3,1})+\frac{k_{b}}{2}(N-\Sigma^{b}_{3,1})-\frac{k_{g}}{2}(N-\Sigma^{g}_{3,1})}
(\xi^{(3)}_{1}\ln \xi^{(3)}_{11}-\xi^{(3)}_{2}\ln
\xi^{(3)}_{2}-\xi^{(3)}_{3}\ln \xi^{(3)}_{3}+\xi^{(3)}_{4}\ln
\xi^{(3)}_{4})\bigg\} \no \\ &&+\frac{1}{4Z_{0}}
\bigg\{e^{\frac{k_{r}}{2}(N-\Sigma^{r}_{4,1})+\frac{k_{b}}{2}(N-\Sigma^{b}_{4,1})+\frac{k_{g}}{2}(N-\Sigma^{g}_{4,1})}
(\xi^{(4)}_{1}\ln \xi^{(4)}_{1}+\xi^{(4)}_{2}\ln
\xi^{(4)}_{2}+\xi^{(4)}_{3}\ln \xi^{(4)}_{3}+\xi^{(4)}_{4}\ln
\xi^{(4)}_{4}) \no
\\&&+e^{-\frac{k_{r}}{2}(N-\Sigma^{r}_{4,1})-\frac{k_{b}}{2}(N-\Sigma^{b}_{4,1})+\frac{k_{g}}{2}(N-\Sigma^{g}_{4,1})}
(-\xi^{(4)}_{1}\ln \xi^{(4)}_{1}+\xi^{(4)}_{2}\ln
\xi^{(4)}_{2}-\xi^{(4)}_{31}\ln \xi^{(4)}_{3}+\xi^{(4)}_{4}\ln
\xi^{(4)}_{4}) \no \\
&&+e^{\frac{k_{r}}{2}(N-\Sigma^{r}_{4,1})-\frac{k_{b}}{2}(N-\Sigma^{b}_{4,1})-\frac{k_{g}}{2}(N-\Sigma^{g}_{4,1})}
(-\xi^{(4)}_{1}\ln \xi^{(4)}_{1}-\xi^{(4)}_{2}\ln
\xi^{(4)}_{2}+\xi^{(4)}_{3}\ln \xi^{(4)}_{3}+\xi^{(4)}_{4}\ln
\xi^{(4)}_{4}) \no \\
&&+e^{-\frac{k_{r}}{2}(N-\Sigma^{r}_{4,1})+\frac{k_{b}}{2}(N-\Sigma^{b}_{4,1})-\frac{k_{g}}{2}(N-\Sigma^{g}_{4,1})}
(\xi^{(4)}_{1}\ln \xi^{(4)}_{1}-\xi^{(4)}_{2}\ln
\xi^{(4)}_{2}-\xi^{(4)}_{3}\ln \xi^{(4)}_{3}+\xi^{(4)}_{4}\ln
\xi^{(4)}_{4})\bigg\}, \no \\ \eea
\end{widetext}

where $S_{cc}(0)$ stands for the topological entanglement of the
color code at zero-temperature, i.e. $S_{cc}(0)=4\ln2$. Notice that
the upper indices of $\xi$'s are related to the respective
bipartitions in Fig.\ref{fig2}(b).


\end{document}